\documentclass[12pt,twoside]{mitthesis}
\usepackage{lgrind}
\usepackage{cmap}
\usepackage[T1]{fontenc}
\usepackage{graphicx}
\usepackage{dcolumn}
\usepackage{bm}
\usepackage{caption}
\usepackage{subcaption}
\usepackage{amsmath}
\usepackage{hyperref}
\DeclareMathOperator*{\argmax}{arg\,max}

\def\all{all}
\ifx\files\all  \else 

\begin{document}

\title{Data-driven extraction of the substructure of quark and gluon jets in proton-proton and heavy-ion collisions}

\author{Yueyang Ying}
\prevdegrees{S.B. Electrical Engineering and Computer Science, Massachusetts Institute of Technology, 2019}
\department{Department of Electrical Engineering and Computer Science}

\degree{Master of Engineering in Electrical Engineering and Computer Science}

\degreemonth{February}
\degreeyear{2022}
\thesisdate{Jan 14, 2022}


\supervisor{Yen-Jie Lee}{Associate Professor of Physics}
\supervisor{Gunther Roland}{Professor of Physics}

\chairman{Katrina LaCurts}{Chair, Master of Engineering Thesis Committee}

\maketitle



\cleardoublepage
\pagestyle{empty}
\setcounter{savepage}{\thepage}
\begin{abstractpage}
%
%
%
The modification of quark- and gluon-initiated jets in the quark-gluon plasma produced in heavy-ion collisions is a long-standing question that has not yet received a definitive answer from experiments. In particular, the size of the modifications in the quark-gluon plasma differs between theoretical models. Therefore a fully data-driven technique is crucial for an unbiased extraction of the quark and gluon jet spectra and substructure. Corroborating past results, I demonstrate the capability of a fully data-driven technique called topic modeling in separating quark and gluon contributions to jet observables. The data-driven topic separation results can further be used to extract jet substructures, such as jet shapes and jet fragmentation function, and their respective QGP modifications. In addition, I propose the use of machine learning constructed observables and demonstrate the potential to increase separability for the input observable. This proof-of-concept study is based on proton-proton and heavy-ion collision events from the PYQUEN generator with statistics accessible in Run 4 of the Large Hadron Collider. These results suggest the potential for an experimental determination of quark- and gluon-jet spectra, their substructures, and their modification in the QGP.
\end{abstractpage}


\cleardoublepage

\section*{Acknowledgments}

This thesis is the culmination of over a year of research that would not have been possible without the mentorship, help, and support of many along the way. In particular, I would like to thank my thesis advisor, Professor Yen-Jie Lee, not only for his insight and guidance throughout this project, but also for being one of the reasons why I pursued physics in the first place. In my time as an undergraduate student, just after I had precariously declared a second major in physics, I took Professor Lee's \textit{Physics III: Vibrations and Waves} course. His charismatic teaching style, along with all the engaging demos, overrode my hesitations at the time about pursuing physics. When I made the decision to return to MIT as a graduate student, I really wanted to work on research at the intersection of machine learning and physics, and Professor Lee made that not only possible, but also a great learning experience. I would also like to thank Professor Gunther Roland and Professor Boleslaw Wyslouch for giving me the opportunity to carry out this research with the group and supporting me throughout this journey.

I am also extremely grateful for the expertise and mentorship of Dr. Yi Chen and Dr. Jasmine Brewer. Their insights were instrumental for the execution of this project and they were always offering guidance when questions arose. With their help, I have been able to learn so much about jets, high-energy physics, and applied machine learning. Thanks to them, I have grown monumentally as a physicist, as a computer scientist, as a presenter, as a writer, and as a person.

I would like to express my thanks to the Relativistic Heavy Ion Group, for funding my studies and introducing me to so many incredibly intelligent, impressive, and inspiring individuals whom I look up to. I am incredibly thankful to have also had the opportunity of a lifetime to work at CERN. I want to thank Dr. Ivan Amos Cali for mentoring me while at CERN, and teaching me all about particle detectors and CMS data. To the other graduate students in the group, thank you all for being my friends and inviting me to all the physics student events, like the LNS student seminars and the physics cookie socials.

A special thanks to my family, for supporting and encouraging me throughout this experience, and to all the amazing friends that I have made in my time at MIT, for making MIT such an incredible place. Finally, a shoutout to MIT COVID-19 procedures, for allowing me to meet the members of the lab in-person, and to my dog, Jessie, for reminding me to take walks outside during the final stretch.


\pagestyle{plain}
\tableofcontents

\chapter{Introduction}
For the first millionth of a second after the Big Bang, the universe was a hot, dense primordial soup, consisting of subatomic particles, before cooling down and forming the ordinary matter. This soup was mostly comprised of quarks (the fundamental building blocks of matter) and gluons (messenger particles that carry the strong force which binds quarks together). This deconfined phase of matter, known as the quark-gluon plasma (QGP), only exists at extremely high temperatures and pressures \cite{big-picture}.

Quantum chromodynamics (QCD) is a component of the widely-accepted Standard Model of physics that focuses on the strong force interactions between quarks and gluons. One motivation behind studying QCD is to better understand the QGP that once dominated the universe. Experiments at ultrarelativistic particle accelerators, such as CERN's Large Hadron Collider (LHC) have been able to recreate the quark-gluon plasma through head-on collisions of heavy ions. However, the QGP exists only for a brief flash of time, less than 10 fm/c, or $O(10^{-23})$ seconds, before cooling and disappearing into ordinary matter, rendering it impossible to probe directly, and thus requiring some alternative way to infer its properties.

In high-energy particle collisions, the collision sometimes produces a high-energy quark or gluon, ``kicked out'' from the nucleus, which then fragments and hadronizes into collimated sprays of particles, known as jets. We reconstruct jets from the particle level by applying a well-defined clustering algorithm, anti-$k_t$ clustering \cite{antikt}, to the individual particles collected in the collision. In heavy-ion collisions, the produced quark-gluon plasma may modify the properties of jets traveling through it. One such modification is jet quenching, a medium-induced parton energy loss phenomena where partons lose energy as they propagate through the QGP due to multiple scattering \cite{jet-quenching, partonenergy, 2005qgp}. Therefore, the phenomena of jet modification makes jets a natural probe to infer qualities of the medium \cite{brewer2020jets}.

One area of interest is to understand how the color charge of partons impacts its traversal through the quark-gluon plasma, as the jet quenching effects of the QGP are different depending on the color charge. Therefore, comprehending the modification of quark- and gluon- jets in the QGP is crucial as it may shed light on such properties. Jets can be modeled in simulation, and such jet quenching models predict certain energy losses for quark and gluon jets, but a data-driven method would be able to directly reveal the quark- and gluon-like jet interaction with the QGP without relying on any theoretical biases \cite{ying}. However, designing a data-driven method to discriminate quark- and gluon-initiated jets in collected jet samples remains an outstanding question \cite{metodiev-topic-of-jets,chien2018probing,jet-substructure,light-quark,Jones:1988ay,Fodor:1989ir,qg-sub,qg-tagging,jet-charge}. Prior work \cite{brewer} demonstrates proof-of-concept success in applying a statistical method known as topic modeling \cite{metodiev-topic-of-jets} to extract quark- and gluon-initiated jet fractions in proton-proton and heavy-ion jet samples.

We apply topic modeling to dijet and $\gamma+$jet samples from PYQUEN \cite{pyquen}, which is a Monte Carlo event generator that simulates medium-induced energy loss of partons in heavy-ion collisions. We operate under the assumption that these input samples are mixtures of the same quark and gluon base distributions. In the topic modeling algorithm, we solve for the quark and gluon fractions in these samples and utilize the results to further extract quark and gluon jet substructure observables in both proton-proton and heavy-ion collisions. The resulting jet substructure observable patterns suggest potential for a fully data-driven mechanism for experimental determination of quark- and gluon-initiated jet spectra and their substructures \cite{ying}. Furthermore, we apply supervised learning techniques to craft new observables that can be input into the unsupervised learning algorithm and demonstrate the potential of machine learning observables to improve topic modeling results.


\chapter{Unsupervised Learning}
Machine learning algorithms can be largely classified into two primary subdomains: supervised learning and unsupervised learning. In supervised learning, one is given a set of data points with corresponding labels, and algorithms are trained to learn the mapping function from the inputs to the outputs. However, in unsupervised learning, one is given a set of unlabeled data, and the goal is to infer some structure or pattern from the given dataset. Unsupervised learning is often used for clustering, anomaly detection, and dimensionality reduction.

In the context of distinguishing quark- and gluon-initiated jets, the samples collected from events at the LHC and RHIC do not contain the truth labels. Thus, we must rely on a data-driven, unsupervised technique to solve this problem. We will discuss the theory behind the algorithm in this section, the simulated data in Section \ref{sec:data}, and the results of the algorithm in Section \ref{sec:results}.

\section{Topic Modeling} \label{sec:topic-modeling}

One well-studied application of unsupervised learning is topic modeling, which is an algorithm used to discover abstract ``topics'' that occur in a collection of text documents \cite{brewer,metodiev-topic-of-jets}. Following the precedence of previous studies, we borrow the term ``topic modeling'' to apply to the problem at hand, where the topics are the type of jet (quark-like/gluon-like), the documents are the histograms of some observable, and the corpus is a collection of histograms \cite{metodiev-topic-of-jets}. The key to topic modeling is the presence of anchor words, which are words unique to a certain topic. In the context of jets, these correspond to anchor bins in the jet observable histogram.

In order to illustrate the concept of topic modeling, we present a small example which we will refer to throughout the chapter. Suppose we have two input distributions, input A and input B, as shown below in Fig. \ref{fig:example-input}. In the case of quark/gluon topic modeling, we assume that these two input distributions are both a combination of the same two unknown base distributions (one quark-like and one gluon-like), or ``topics''. The goal of the algorithm is to derive these distributions.

\begin{figure}[htp]
    \centering
    \includegraphics[width=.55\textwidth]{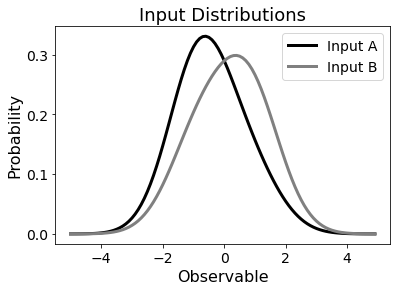}
    \caption{Example of two input distributions that might be used for topic modeling.}
    \label{fig:example-input}
\end{figure}

\begin{figure}[htp]
    \centering
    \begin{subfigure}{.49\textwidth}
        \centering
        \includegraphics[width=\textwidth]{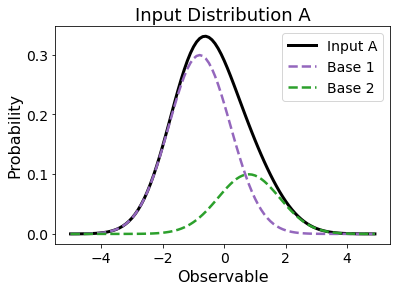}
        \caption{Base distributions of input A}
        \label{fig:input-A}
    \end{subfigure}
    \begin{subfigure}{.49\textwidth}
        \centering
        \includegraphics[width=\textwidth]{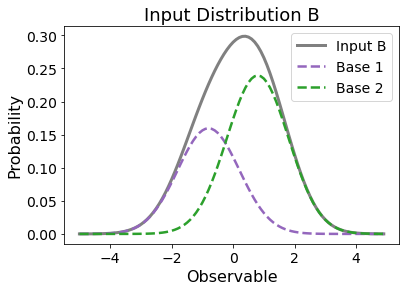}
        \caption{Base distributions of input B}
        \label{fig:input-B}
    \end{subfigure}
    \caption{The input distributions are the result of superimposed base distributions.}
    \label{fig:input-mixtures}
\end{figure}

Similarly, jet samples collected from colliders are mixtures of these jet topics. Each jet observable histogram is a mixture of the two underlying quark/gluon base distributions. Mathematically, we can represent the input histograms as
\begin{equation} \label{eq:mix}
    p^{(s)}_O(x) = f^{(s)} p_1(x) + (1-f^{(s)}) p_2(x)
\end{equation}
where $p^{(s)}_O(x)$ represents the density distribution in sample $s$ with respect to some observable $O$, $f^{(s)}$ represents the fraction of topic 1 in sample $s$, $1-f^{(s)}$ represents the fraction of topic 1 in sample $s$, and $p_i(x)$ represent the base distributions, which correspond to the jet topics. Fig. \ref{fig:input-mixtures} illustrates the contributions from the underlying base distributions, which superimpose to obtain the example input distribution.

However, in Eq. \ref{eq:mix}, there remains some ambiguity as there are infinitely many ways to define $p_1$ and $p_2$ and modify $f^{(s)}$ accordingly such that the equation remains true. In order to resolve this, we use the \texttt{DEMIX} algorithm \cite{katzsamuels2019decontamination}, which breaks this ambiguity by choosing unique base distributions for $p_1$ and $p_2$, which we represent using $b_1$ and $b_2$, respectively.

\begin{figure}[htp]
    \centering
    \includegraphics[width=.55\textwidth]{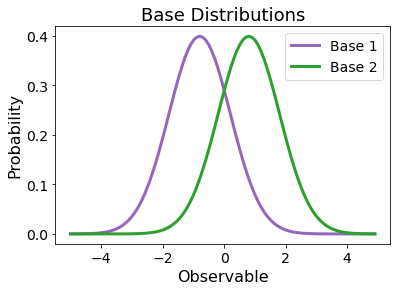}
    \caption{The base distributions underlying the example input.}
    \label{fig:example-base}
\end{figure}

\texttt{DEMIX} results in the \textit{mutually irreducible} \cite{blanchard2016classification} underlying distributions, $b_1$ and $b_2$. This is synonymous with requiring the presence of anchor bins, or requiring each underlying distribution to be not be a mixture of one another plus another distribution. In other words, we cannot write $b_1(x) = cb_2(x) + (1-c)F$, and vice versa, where $F$ is some probability distribution and $0 < c \leq 1$. This also implies that $\lim_{x \rightarrow \infty} b_1(x)/b_2(x) = 0$ and $\lim_{x \rightarrow -\infty} b_2(x)/b_1(x) = 0$ \footnote{Depending on how we define $b_1(x)$ and $b_2(x)$, the limits may be reversed. That is, we may find $\lim_{x \rightarrow -\infty} b_1(x)/b_2(x) = 0$ and $\lim_{x \rightarrow \infty} b_2(x)/b_1(x) = 0$}. The two mutually irreducible base distributions underlying our example input distributions are shown in Fig. \ref{fig:example-base}. The goal of this data-driven algorithm is thus to obtain these two base distributions, along with the mixture fractions of each topic, $f$ and $1-f$, in each input sample.

It is worth noting that \texttt{DEMIX} requires the input distributions to be sample independent and contain different base purities, and guarantees that the two resulting base distributions are mutually irreducible. Ref. \cite{komiske} defines the \textit{operational definition} quark and gluon categories as the mutually irreducible underlying distributions given two mixed jet samples:
\\\\
\textbf{Quark/gluon jet definition (operational).} \cite{komiske} \textit{Given two samples \(A\) and \(B\) of QCD jets at a fixed \(p_T\) obtained by a suitable jet-finding procedure, taking \(A\) to be ``quark-enriched'' compared to \(B\), and a jet substructure feature space \(\mathcal{O}\), the quark and gluon jet distributions are defined to be:}
$$
p_q(\mathcal{O}) \equiv \frac{p_{A}\mathcal{O} - \kappa_{AB} p_{B}(\mathcal{O})}{1-\kappa_{AB}}, \;\;\;\;\;\;\;\;
p_g(\mathcal{O}) \equiv \frac{p_{B}\mathcal{O} - \kappa_{BA} p_{A}(\mathcal{O})}{1-\kappa_{BA}}
$$
\textit{where $\kappa_{AB}$, $\kappa_{BA}$, $p_{A}(\mathcal{O})$, $p_{B}(\mathcal{O})$ are directly attainable from \(A\) and \(B\).}
\\

Therefore, according to this definition, the topics resulting from \texttt{DEMIX} will correspond to the quark- and gluon-like jet distributions. We choose to use constituent multiplicity (number of constituent particles in a given jet) as the input observable because it has been shown to be mutually irreducible in the high-energy limit \cite{metodiev-topic-of-jets}.

To extract the base distributions, we need to find the reducibility factor $\kappa$, which represents the largest amount one distribution can be subtracted from the other, such that all bins remain non-negative:
\begin{equation} \label{kappa}
    \kappa_{ij} = \inf_x \frac{p_i(x)}{p_j(x)}
\end{equation}

Then, with $b_1$ and $b_2$ corresponding to the distributions for topic 1 and topic 2, respectively, and $p_A$ and $p_B$ corresponding to the distributions for the two input samples, we are able to derive the base distributions using the following equations:
\begin{equation} \label{eq:topics}
    \begin{split}
        b_1(x) = \frac{p_A(x) - \kappa_{AB}p_B(x)}{1-\kappa_{AB}},\\
        b_2(x) = \frac{p_B(x) - \kappa_{BA}p_A(x)}{1-\kappa_{BA}}
    \end{split}
\end{equation}

However, in the context of collider physics, $p_A$ and $p_B$ are finite histograms with $n$ bins. For a finite-sampled distribution, the $\kappa$ values are extracted from the tails, where the statistics are low. In order to combat this, we can leverage the additional statistics from the middle of the distribution by fitting a sum of skew-normal distributions to the sampled histograms, expressed as
\begin{equation}
    f_N(x; \alpha_i, \theta) = \sum_{k=1}^N \alpha_{i,k} \text{SN}(x; \mu_k, \sigma_k, s_k)
\end{equation}
where $\text{SN}(x; \mu_k, \sigma_k, s_k)$ represents a skew-normal distribution with parameters $\mu_k$, $\sigma_k$, and $s_k$. While the mixture fractions, $\alpha_i$, are unique to the input histograms, $ \mu_k, \sigma_k, s_k$ are shared between the two. For generality, we use $N=4$, such that we have 18 fit parameters, representing $\alpha_A$, $\alpha_B$, and $\theta$ \cite{brewer}.

Let $i$ represent the $i$th jet sample, $j$ represent the bin index of the corresponding histogram, such that $n_{i,j} = n_iy_{i,j}$ represents the count in bin $j$ of the $i$th sample, where $y_{i,j}$ is the probability density value of the bin. We assume that the count follows a Poisson distribution with mean value $\mu_{i,j}(\alpha_i, \theta) = f(x_{i,j}; \alpha_i,\theta)$. The best-fit parameters and the corresponding uncertainties can thus be captured by the Poisson-likelihood chi-square function \cite{BAKER1984437, chi-square}:
\begin{equation}
    \ln{\frac{C}{p}} = \sum_{i,j} n_i \left[ \mu_{i,j}(\alpha_i, \theta) - y_{i,j} + y_{i,j} \ln{\frac{y_{i,j}}{\mu_{i,j}(\alpha_i, \theta)}} \right]
\end{equation}
where $p$ represents the probability and $C$ represents a constant scaling factor, introduced  to simply cancel out a term that comes from taking the log of the Poisson distribution \cite{brewer, BAKER1984437}.

In order to extract the parameter values and uncertainties from the likelihood function, we use Markov chain Monte Carlo (MCMC) \cite{emcee}. We obtain initial estimates of the parameter values by running a simultaneous least-squares fit. For the results shown in this paper, we use 100 MCMC walkers, initialized using the least-squares parameters, and run for 35,000 samples using a burn-in of 30,000 samples.

\begin{figure}[htp]
    \centering
    \begin{subfigure}{.49\textwidth}
        \centering
        \includegraphics[width=\textwidth]{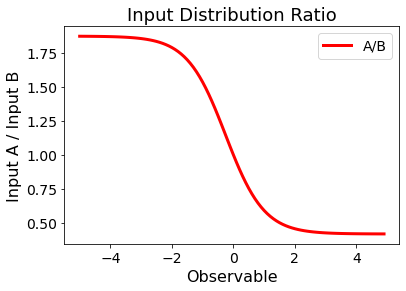}
        \caption{}
        \label{fig:ab}
    \end{subfigure}
    \begin{subfigure}{.49\textwidth}
        \centering
        \includegraphics[width=\textwidth]{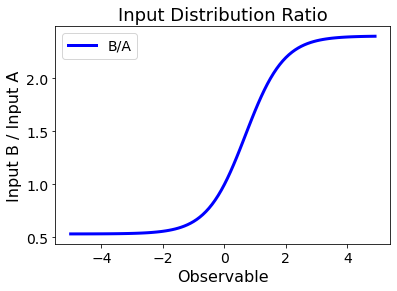}
        \caption{}
        \label{fig:ba}
    \end{subfigure}
    \caption{}
    \label{fig:ratios}
\end{figure}

Once we have the MCMC fits for the input distributions, we can take ratios between the inputs' skew-normal fits. Fig. \ref{fig:ratios} displays a conceptual image of what the input ratios look like using our example distributions. Here, the reducibility factor for input A / input B, $\kappa_{AB}$, would be extracted from the right tail of Fig. \ref{fig:ab}, and similarly, $\kappa_{BA}$, would be extracted from the left tail of Fig. \ref{fig:ba}, as that is where the minimum of the curve is located.

However, in the above example, there is only a single curve, as the example inputs are well-defined. In reality, when we obtain results from the MCMC, there will be multiple curves, each corresponding to a fit from the walkers, and each with a different minima, or $\kappa$ value. In order to extract the topics and calculate the corresponding uncertainty using Eq. \ref{eq:topics}, we sample for $\kappa_{AB}$ and $\kappa_{BA}$ from the fits, and calculate the mean and standard deviation.
\chapter{Simulated Collision Events} \label{sec:data}
This proof-of-concept study is based on proton-proton and heavy-ion collision events from the PYQUEN generator \cite{pyquen} with statistics accessible in Run 4 of the Large Hadron Collider. The PYQUEN event generator simulates of rescattering, and radiative and collisional energy loss of partons in the QGP in heavy-ion collisions \cite{pyquen}. The input distributions are photon-jet ($\gamma+$jet) samples and dijet samples. We choose these because at Large Hadron Collider energies, $\gamma+$jet and dijets have different quark and gluon jet fractions. The modified jets are not embedded in thermal background.
      
The proton-proton and heavy-ion events are produced using $\hat{p_T}>80$ GeV, where $\hat{p_T}$ is the hard scattering scale, and an impact parameter range corresponding to 0-10\% centrality. We use \textsc{FastJet} 3.3.0 \cite{fastjet,kt} to reconstruct anti-$k_t$ jets with radius $R=0.4$ \cite{antikt}. For this analysis, in the $\gamma+$jet samples, we select the leading jet in the opposite direction azimuthally ($|\Delta \phi| > \pi/2$) to the high-momentum photon, and in the dijet samples, we select the two jets with largest transverse momenta.

We only include jets with $80 < p_T < 100$ GeV and we impose a cut of $|\eta| < 1$. In addition, there was a low multiplicity peak in the $\gamma+$jet sample, composed of particles that surrounded a high-momentum photon. Therefore, we also imposed a photon $p_T$ ratio cut by determining the highest momentum photon within the jet, $\gamma_h$, and removing any jets where $p_{T}^{(\gamma_h)}/p_{T}^{(\text{jet})} > 0.8$, which resolved the low multiplicity peak.

Lastly, one critical piece of information to assess the performance of the topic modeling algorithm is the Monte Carlo truth labels for quark-like and gluon-like jets. Although these labels are not quite well-defined, in order to determine such labels, we build an approximation by comparing angular distance between the selected jet and the two outgoing matrix elements in the simulation. For $\gamma+$jet, we simply label the jet under the outgoing matrix element that is not the photon. For dijets, we match the jet to the outgoing matrix element with the smallest angular distance, $\Delta R = \sqrt{(\Delta \eta)^2 + (\Delta \phi)^2}$, from the jet.

Furthermore, in the MC truth labels, we only include samples where $\Delta R < 0.4$ between the matrix element and the jet. In the pp $\gamma+$jet sample, we utilize 97\% of the jets for the quark/gluon truth label, and in the pp dijet sample, we utilize 92\% of jets for the quark/gluon truth label. In the PbPb sample, we utilize 95\% and 89\% for $\gamma+$jet and dijet quark/gluon truth labels, respectively. Ultimately, this means that the MC truth in any of the results should not be taken as the absolute truth, but rather just an approximation \cite{ying}.
\chapter{Topic Modeling Results}
\label{sec:results}

\section{MCMC and Resulting Topics}

\begin{figure*}[htp]
    \centering
    \begin{subfigure}{0.45\textwidth}
        \centering
        \includegraphics[width=\textwidth]{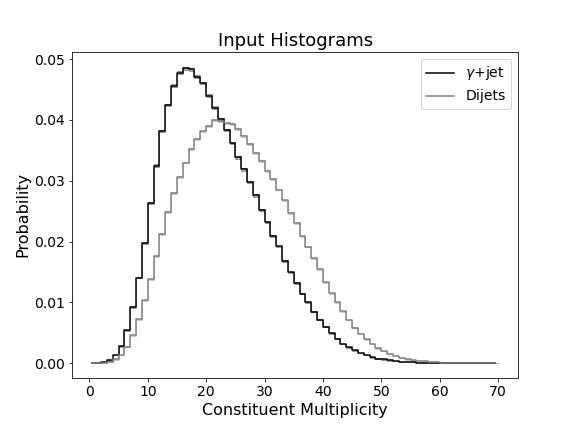}
        \caption{Proton-proton input distributions}
        \label{fig:pp-input}
    \end{subfigure}
    \begin{subfigure}{0.45\textwidth}
        \centering
        \includegraphics[width=\textwidth]{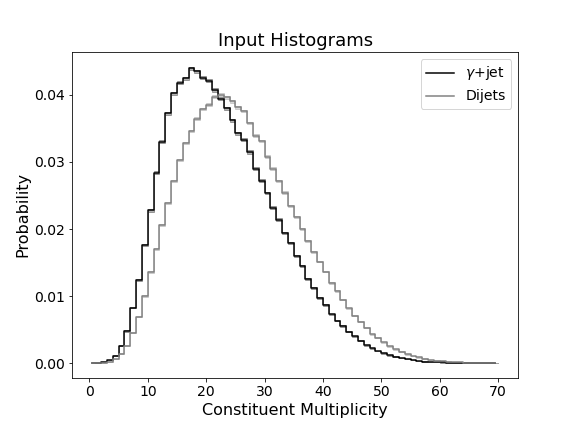}
        \caption{Heavy-ion input distributions}
        \label{fig:hi-input}
    \end{subfigure}
    \caption{PYQUEN-generated proton-proton (left) and heavy-ion (right) normalized constituent multiplicity input distributions for $\gamma+$jet and dijets.}
    \label{fig:inputs}
\end{figure*}

In this section, we demonstrate the results of the topic modeling algorithm on the PYQUEN data. Since previous work \cite{metodiev-topic-of-jets} has demonstrated that constituent multiplicity approximately satisfies quark-gluon mutual irreducibility, we use this observable as input into the topic modeling machinery as a starting point. The input distributions are shown in Fig. \ref{fig:inputs}. Fig. \ref{fig:pp-input} shows the $\gamma+$jet and dijets histograms for proton-proton collision events and Fig. \ref{fig:hi-input} shows the input histograms for the heavy-ion events.

\begin{figure*}[htp]
    \centering
    \begin{subfigure}{\textwidth}
        \centering
        \includegraphics[width=\textwidth]{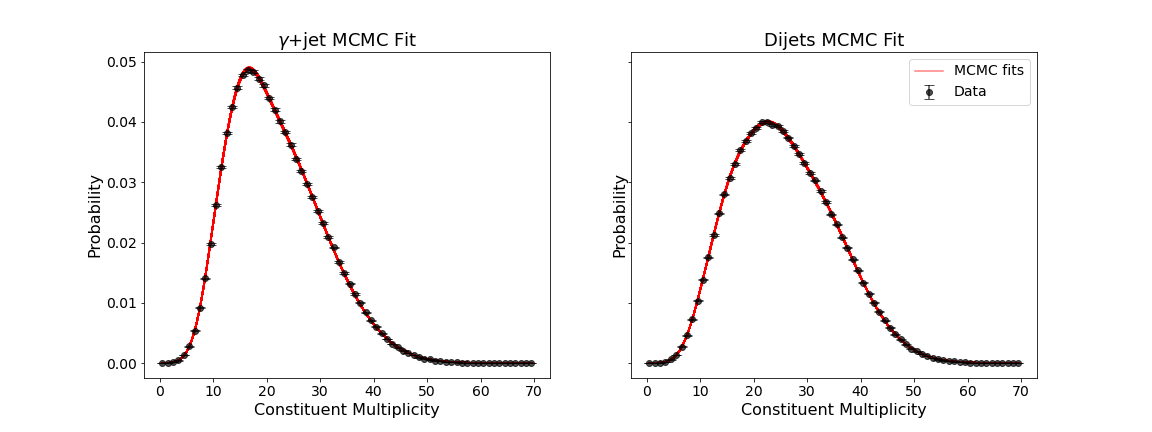}
        \caption{Proton-proton MCMC fits}
        \label{fig:pp-mcmc}
    \end{subfigure}
    \begin{subfigure}{\textwidth}
        \centering
        \includegraphics[width=\textwidth]{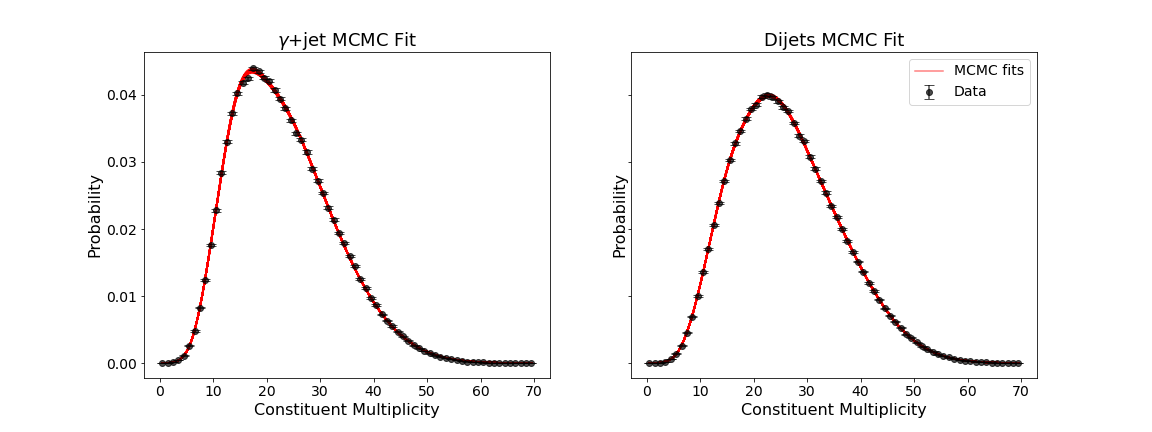}
        \caption{Heavy-ion MCMC fits}
        \label{fig:hi-mcmc}
    \end{subfigure}
    \caption{MCMC fits for proton-proton (top) and heavy-ion (bottom) input distributions. Each red line represents a single fit (set of parameters) dictated by the MCMC walkers.}
    \label{fig:mcmc}
\end{figure*}

The Markov chain Monte Carlo fits for both the $\gamma+$jet and dijets histograms are shown in Fig. \ref{fig:mcmc}. Each red line represents one set of parameters from the MCMC for the following equation \cite{brewer}:
\begin{equation}
    f_4(x; \alpha_i, \theta) = \sum_{k=1}^4 \alpha_{i,k} \text{SN}(x; \mu_k, \sigma_k, s_k)
\end{equation}

\begin{figure*}[htp]
    \centering
    \begin{subfigure}{\textwidth}
        \centering
        \includegraphics[width=\textwidth]{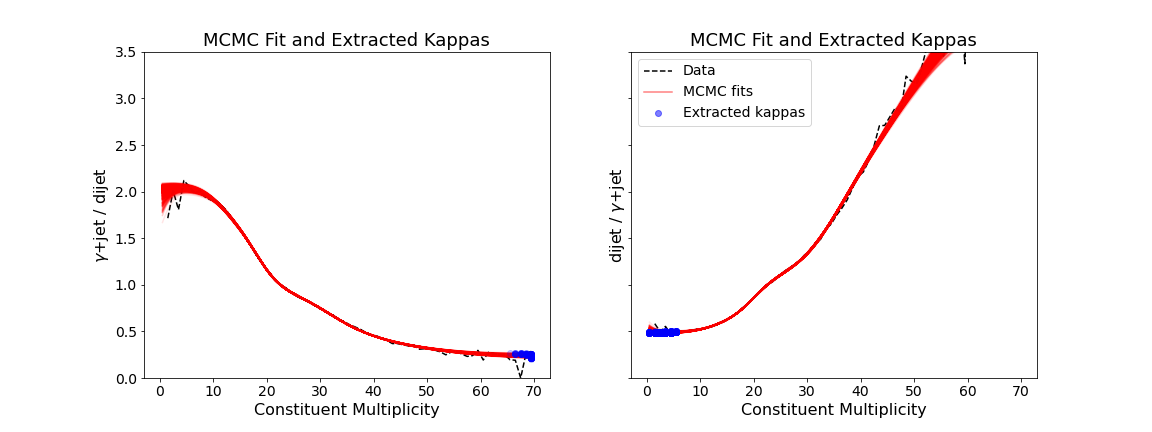}
        \caption{Proton-proton kappa extraction}
        \label{fig:pp-kappas}
    \end{subfigure}
    \begin{subfigure}{\textwidth}
        \centering
        \includegraphics[width=\textwidth]{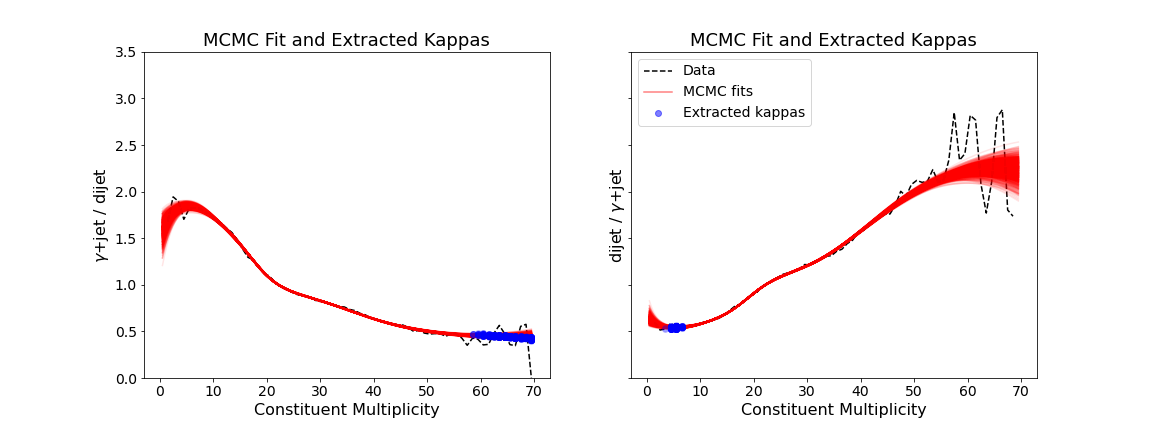}
        \caption{Heavy-ion kappa extraction}
        \label{fig:hi-kappas}
    \end{subfigure}
    \caption{Ratio of $\gamma+$jet / dijet (left) and dijet / $\gamma+$jet (right) MCMC fits for proton-proton (top) and heavy-ion (bottom). These red lines represent the ratios of the MCMC fits from Fig. $\ref{fig:mcmc}$. The real value from data is shown in black, and the sampled $\kappa$ values are shown in blue.}
    \label{fig:kappas}
\end{figure*}

\begin{figure*}[htp]
    \centering
    \begin{subfigure}{0.45\textwidth}
        \centering
        \includegraphics[width=\textwidth]{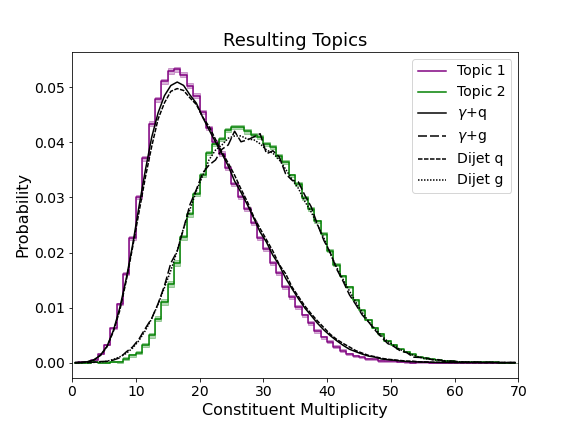}
        \caption{Proton-proton topic separation results}
        \label{fig:pp-topics}
    \end{subfigure}
    \begin{subfigure}{0.45\textwidth}
        \centering
        \includegraphics[width=\textwidth]{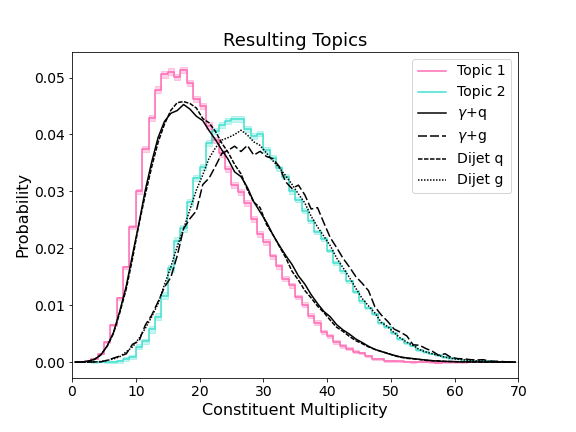}
        \caption{Heavy-ion topic separation results}
        \label{fig:hi-topics}
    \end{subfigure}
    \caption{Proton-proton (left) and heavy-ion (right) topic extraction results, displaying the resulting topics in comparison to the gluon and quark MC truth labels.}
    \label{fig:topics}
\end{figure*}

These MCMC fits are then used to find the ratio between the $\gamma+$jet and dijet distribution, or $f_4^{\gamma+\text{jet}}(x; \alpha_i, \theta) / f_4^{\text{dijet}}(x; \alpha_i, \theta)$, and vice versa, as shown in Fig. \ref{fig:kappas}. As demonstrated in Eq. \ref{kappa}, the desired $\kappa$ value to extract is the infimum of the ratio of the input distributions, which in the finite case, becomes the minimum of the MCMC fit ratios. The extracted $\kappa$ values are shown in blue in Fig. \ref{fig:kappas}. Following Eq. \ref{eq:topics}, using the extracted $\kappa$ values, we can thus calculate the base distributions (topics) given the input histograms. The resulting topics for both the proton-proton input, as well as the heavy-ion input, are shown in Fig. \ref{fig:topics}.

In general, the extracted topics correspond fairly well to the MC truth distributions, with topic 1 being quark-like and topic 2 being gluon-like. However, the topic-truth matching appears slightly better for proton-proton than for heavy-ion, though there appears to be some disparity between topic 1 and the quark-like truth, which is exacerbated in the heavy-ion sample. It is worth noting that previous work \cite{brewer} has demonstrated similar success in extraction of the topics using \texttt{DEMIX} in both pp and PbPb samples, but using a different model (JEWEL). Thus, it is non-trivial that we have corroborated such findings by reproducing similar results in PYQUEN \cite{ying}.

\section{Substructure Observable Extraction}
While the data-driven determination of quark- and gluon-like jet fractions in a sample is significant, applying these results to extract quark and gluon jet substructure allows for deeper insight into the modification of quark and gluon jets in the quark gluon plasma. In this section, we demonstrate the application of the topic modeling algorithm to find jet observable distributions corresponding to the topics, and compare these to the MC truth quark and gluon jet observable distributions. Namely, we will take a look at jet shape, jet fragmentation, jet mass, and jet splitting function. We also utilize these results to determine the modification of quark and gluon jet observable by taking the ratio of jet observable between the heavy-ion sample and the proton-proton sample \cite{ying}.

\subsection{Jet Shape Extraction}
\begin{figure}[htp]
    \centering
    \begin{subfigure}{0.52\textwidth}
        \centering
        \includegraphics[width=\textwidth]{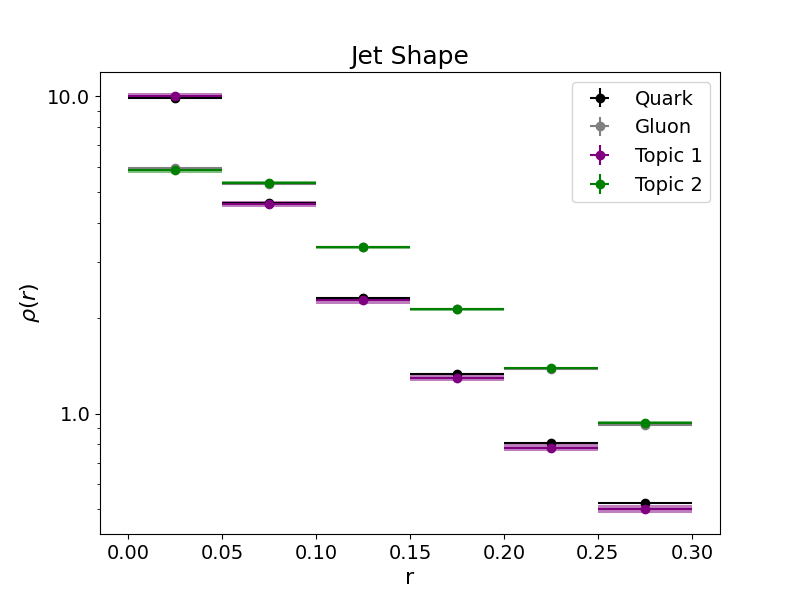}
        \caption{Proton-proton jet shape}
        \label{fig:pp-jet-shape}
    \end{subfigure}
    \begin{subfigure}{0.52\textwidth}
        \centering
        \includegraphics[width=\textwidth]{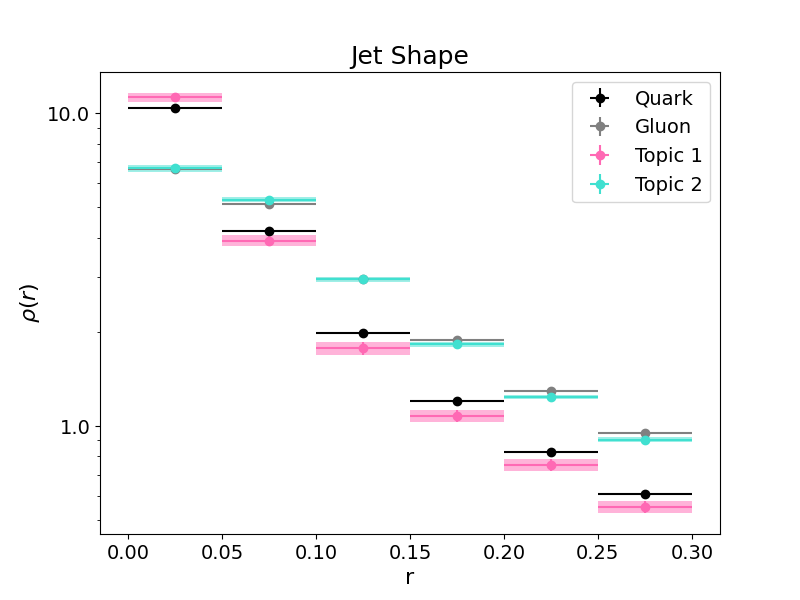}
        \caption{Heavy-ion jet shape}
        \label{fig:hi-jet-shape}
    \end{subfigure}
    \begin{subfigure}{0.52\textwidth}
        \centering
        \includegraphics[width=\textwidth]{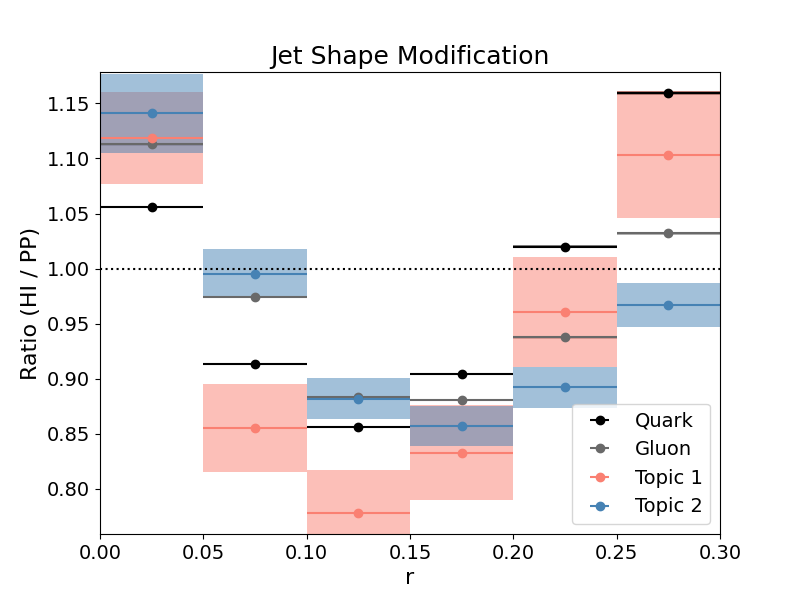}
        \caption{Jet shape modification ratio}
        \label{fig:mod-shape}
    \end{subfigure}
    \caption{Proton-proton (top) and heavy-ion (middle) jet shape extraction using topic modeling results from Fig. \ref{fig:topics}. The modification of jet shape in the quark-gluon plasma is shown as the ratio between heavy-ion and proton-proton jet shape (bottom).}
    \label{fig:jet-shape}
    \end{figure}

We first apply our findings to derive the topics' jet shape. Jet shape describes the jet transverse momentum distribution as a function of radial distance from the jet axis, and can be described by the following equation
\begin{equation}
    \rho(r) = \frac{1}{r_b - r_a}\frac{1}{N_{jet}} \sum_{jets}\frac{\sum_{tracks\in[r_a, r_b)}p_T^{track}}{p_T^{jet}}
\end{equation}
Here, $r$ corresponds to the radial distance from the jet axis, and $r_a$, $r_b$ correspond to the inner and outer radii of the given annulus \cite{jet-shape}. Each annulus corresponds to a bin in the jet shape plot, where $r_a$ is the left edge of the bin and $r_b$ is the right edge of the bin.

In order to obtain the jet shape using our topic modeling results, we can simply perform a linear combination using the extracted $\kappa$ values for each bin in the jet shape:
\begin{equation}
\begin{split}
    \rho_{1}(r) = \frac{\rho_{\gamma+\text{jet}}(r) - \kappa_{AB} \rho_{\text{dijets}}(r)}{1 - \kappa_{AB}},\\
    \rho_{2}(r) = \frac{\rho_{\text{dijets}}(r) - \kappa_{BA} \rho_{\gamma+\text{jet}}(r)}{1 - \kappa_{BA}}
\end{split}
\end{equation}
Here, $\rho_{\gamma+\text{jet}}(r)$ and $\rho_{\text{dijets}}(r)$ are the jet shapes for the $\gamma+\text{jet}$ and $\text{dijets}$, respectively.

The resulting topic jet shapes, along with the MC truth quark and gluon jet shapes, are shown in Fig. \ref{fig:jet-shape}. The gluon jet shape and the topic 2 jet shape appear to be in very good agreement in both the proton-proton and heavy-ion samples. On the other hand, the quark jet shape and the topic 1 jet shape demonstrate similar trends in pattern, but it appears that the topic 1 jets tend to be slightly narrower than the quark-like jets in the heavy-ion sample.

We then computed the jet shape modification (ratio between heavy ion jet shape and proton-proton jet shape), shown in Fig. \ref{fig:mod-shape}. In the jet shape modification plot, while the extracted topic ratios are able to roughly match the quark and gluon ratios in terms of general trend, there does appear to be a disparity up to 8\% between the topic and the MC truth in some bins.

\subsection{Jet Fragmentation Function Extraction}
The topic modeling results can also be used to extract the quark and gluon jet fragmentation function. The jet fragmentation function represents the longitudinal momentum distribution of the tracks inside a jet, and can be expressed by the following equation
\begin{equation}
    D(\xi) = \frac{1}{N_{jet}}\frac{dN_{track}}{d\xi}
\end{equation}
Here, $N_{jet}$ is the total number of jets, $N_{track}$ is the number of tracks in a jet, and $\xi = \ln{(1/z)}$, where $z$ is the longitudinal momentum fraction, defined as
\begin{equation} 
    z = \frac{p_T\cos{\Delta R}}{p_T^{jet}} = \frac{p_T}{p_T^{jet}}\cos{\sqrt{(\Delta \eta)^2 + (\Delta \phi)^2}}
\end{equation}
Here, $p_T^{jet}$ is the transverse momentum of the jet relative to the beam direction, $p_T$ is the transverse momentum of a charged particle in the jet, and $\Delta \eta$ and $\Delta \phi$ are measures of distance between the particle and E-scheme jet axis in pseudorapidity and azimuth \cite{jet-frag}.

In the jet fragmentation function, we can also compute each bin of the topics using a linear combination as shown below.
\begin{equation}
\begin{split}
    D_{1}(\xi) = \frac{D_{\gamma+\text{jet}}(\xi) - \kappa_{AB} D_{\text{dijets}}(\xi)}{1 - \kappa_{AB}},\\
    D_{2}(\xi) = \frac{D_{\text{dijets}}(\xi) - \kappa_{BA} D_{\gamma+\text{jet}}(\xi)}{1 - \kappa_{BA}}
\end{split}
\end{equation}

One note here is that previous, in calculating the jet shape, the formula includes a normalization factor. In the jet fragmentation function, the histogram, $D(\xi)$, is normalized by the total number of jets, such that the integral of the histogram over $\xi$ represents the average number of charged particles per jet. Therefore, rather than normalize for density, we take the direct combination of the per-jet quantities in each bin, since we want the output to be a per-jet quantity. To demonstrate this, by definition,
\begin{equation}
\begin{split}
    D_{\text{dijets}}(\xi) = f_d D_1(\xi) + (1-f_d) D_2(\xi)\\
    D_{\gamma+\text{jet}}(\xi) = f_\gamma D_1(\xi) + (1-f_\gamma) D_2(\xi)
\end{split}
\end{equation}

Now, after we integrate, we arrive at the following set of equations:
\begin{equation}
\begin{split}
    N_{\text{tracks}}^{(\text{dijets})} = f_d N_{\text{tracks}}^{(\text{topic 1})} + (1-f_d) N_{\text{tracks}}^{(\text{topic 2})}\\
    N_{tracks}^{(\gamma+\text{jet})} = f_\gamma N_{\text{tracks}}^{(\text{topic 1})} + (1-f_\gamma) N_{\text{tracks}}^{(\text{topic 2})}
\end{split}
\end{equation}
which demonstrates that the average number of tracks in dijets (or $\gamma+$jets) equates to the weighted average of topic 1's average number of tracks and topic 2's average number of tracks. Therefore, rather than include any normalization, we directly apply $\kappa$ to the dijet and $\gamma+jet$ jet fragmentation values in order to solve for jet fragmentation of topic 1 and topic 2.

\begin{figure}[htp]
    \centering
    \begin{subfigure}{0.52\textwidth}
        \centering
        \includegraphics[width=\textwidth]{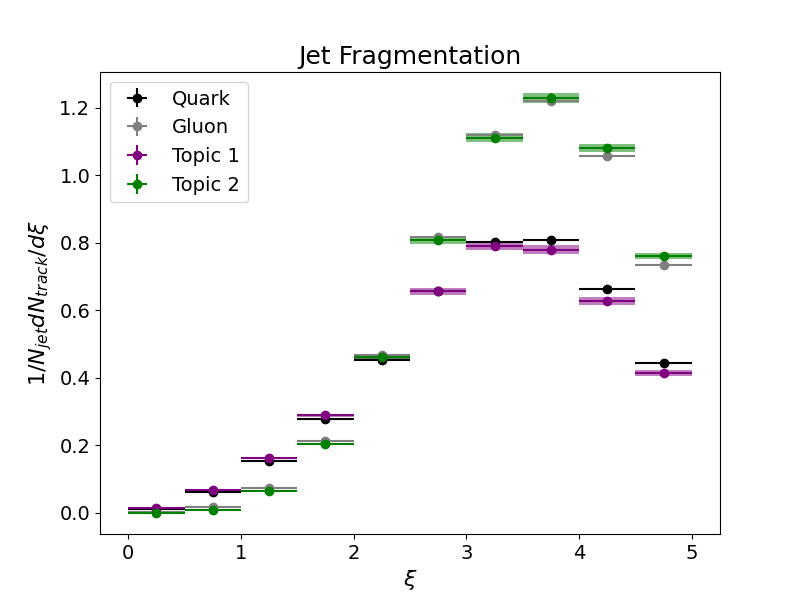}
        \caption{Proton-proton}
        \label{fig:pp-nolog}
    \end{subfigure}
    \begin{subfigure}{0.52\textwidth}
        \centering
        \includegraphics[width=\textwidth]{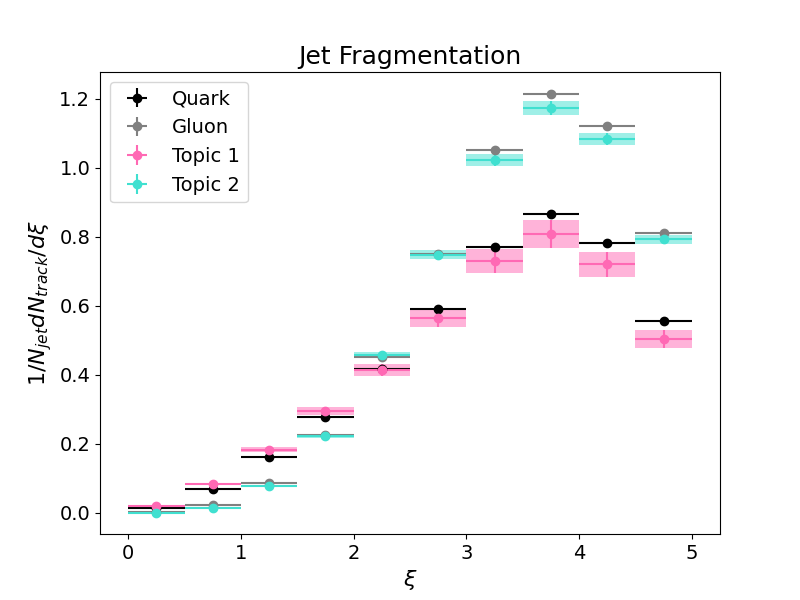}
        \caption{Heavy-ion}
        \label{fig:pbpb-nolog}
    \end{subfigure}
    \begin{subfigure}{0.52\textwidth}
        \centering
        \includegraphics[width=\textwidth]{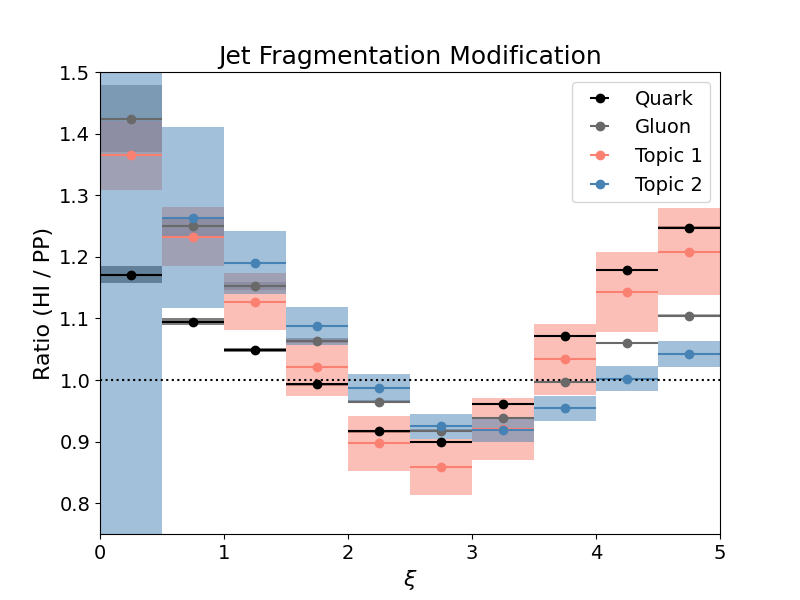}
        \caption{Jet fragmentation modification ratio}
        \label{fig:mod-frag}
    \end{subfigure}
    \caption{Extracting jet fragmentation for proton-proton (top) and heavy-ion (middle) collision using topic modeling results from Fig. \ref{fig:topics}. The jet fragmentation modification, represented by the ratio between PbPb and pp jet fragmentation, is also shown (bottom) for the topics.}
    \label{fig:jet-frag}
\end{figure}

Fig. \ref{fig:jet-frag} shows the extracted topic 1 and topic 2 jet fragmentation, as well as the MC truth quark and gluon jet fragmentation, in proton-proton and heavy-ion collisions. In both samples, the topic 1 agreement with the quark fragmentation function and the topic 2 agreement with the gluon fragmentation function are extremely similar.

We observe that at high track $p_T$, such as when $\xi < 2$, it appears that our topic modeling fractions tend to underestimate the gluon fragmentation, but overestimate the quark fragmentation. At $2 < \xi < 3$, the fragmentation function of the topics matches their respective MC truth fragmentation functions to within the error bound. At higher $\xi$, it appears that the extracted topic fragmentation functions underestimate the values for the quark fragmentation and overestimate the values for the gluon fragmentation in the proton-proton sample. However, both topics appear to underestimate the MC truths in the heavy-ion sample.

In the jet fragmentation modification plot in Fig. \ref{fig:mod-frag}, we see a similar rough matching in the middle, such as between $1 < \xi < 4$. However, at $\xi < 1$, there appears to be a large discrepancy between quark and topic 1, as well as gluon and topic 2, assuming the trend that topic 1 is more quark-like and topic 2 is more gluon-like holds. We also note that while there is a large discrepancy, the errors in these two bins are also quite large, due to a small value in the corresponding jet fragmentation bins and thus large relative error. On the other hand, at $\xi > 4 $, there appears to be a generally close match between the quark and topic 1 ratios, but the ratio for the gluon is outside the error range for the topic 2 ratio. This is interesting, as previously, we had seen a closer match between gluon and topic 2, in comparison to the match between quark and topic 1.

\subsection{Jet Mass Extraction}

In addition to jet shape and jet fragmentation, which measure the distribution of energy in specified areas of the jet cone, we also perform this exercise on two additional per-jet substructure observables: jet mass and jet splitting function.

\begin{figure}[htp]
    \centering
    \begin{subfigure}{0.52\textwidth}
        \centering
        \includegraphics[width=\textwidth]{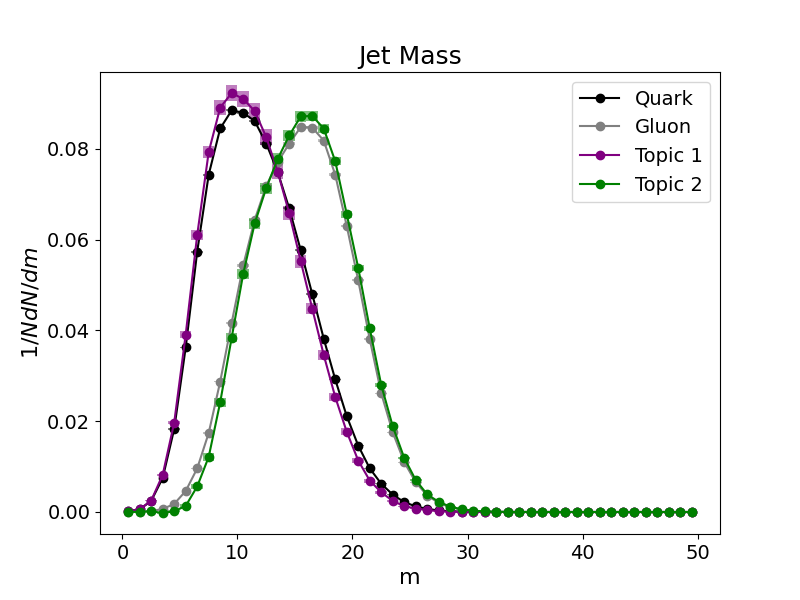}
        \caption{Proton-proton jet mass}
        \label{fig:pp-jet-mass}
    \end{subfigure}
    \begin{subfigure}{0.52\textwidth}
        \centering
        \includegraphics[width=\textwidth]{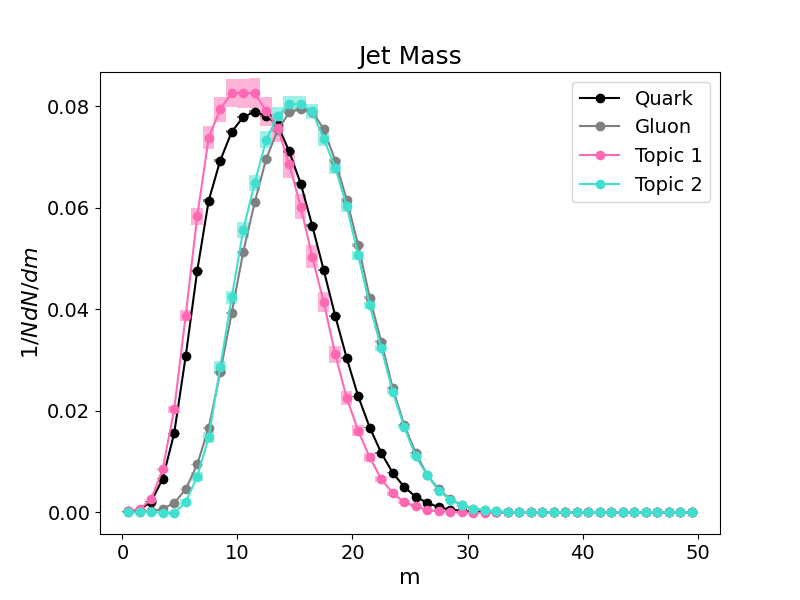}
        \caption{Heavy-ion jet mass}
        \label{fig:hi-jet-mass}
    \end{subfigure}
    \begin{subfigure}{0.52\textwidth}
        \centering
        \includegraphics[width=\textwidth]{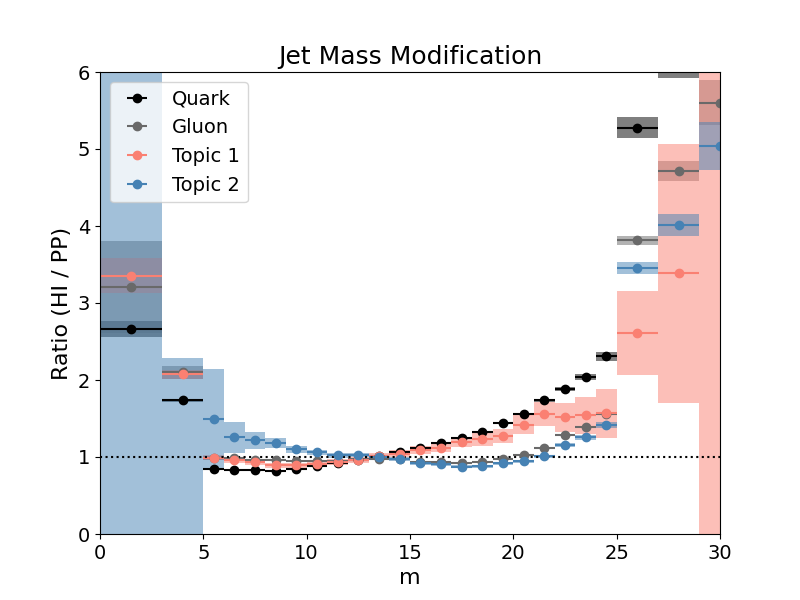}
        \caption{Jet mass modification ratio}
        \label{fig:mod-mass}
    \end{subfigure}
    \caption{Topic modeling results for proton-proton (top) and heavy-ion (middle) jet mass, compared to MC truth. Modification of the jet mass in the QGP, as defined by the ratio of the heavy-ion and proton-proton jet mass spectra is shown as well (bottom).}
    \label{fig:jet-mass}
\end{figure}

The jet mass is calculated from the jet four-momentum, the total four-momentum of all the constituents in the jet, and is expressed as $m = \sqrt{(E+p_z)*(E-p_z)-p_T^2}$, where $E$ is the jet energy, $p_z$ is the longitudinal momentum of the jet, and $p_T$ is the transverse momentum of the jet. To extract the topics' jet mass histogram using our topic modeling results, we perform a linear combination on the normalized jet mass input histograms, $H_{\gamma+\text{jet}}(m)$ and $H_{\text{dijets}}(m)$, using the extracted $\kappa$ values:
\begin{equation}
\begin{split}
    H_{1}(m) = \frac{H_{\gamma+\text{jet}}(m) - \kappa_{AB} H_{\text{dijets}}(m)}{1 - \kappa_{AB}},\\
    H_{2}(m) = \frac{H_{\text{dijets}}(m) - \kappa_{BA} H_{\gamma+\text{jet}}(m)}{1 - \kappa_{BA}}
\end{split}
\end{equation}

The resulting topic 1 and topic 2 normalized jet mass histograms for pp and PbPb are shown in Fig. \ref{fig:jet-mass}, along with the normalized jet mass histograms for the MC-labelled quark and gluon samples. The ratio between the PbPb and pp jet mass histogram bins is shown as well.

For both proton-proton and heavy-ion, the results corroborate previous jet shape and jet fragmentation results that demonstrate topic 1's correspondence to the quark distribution and topic 2's correspondence to the gluon distribution. While it appears that the extracted topics are slightly narrower than the quark and gluon ``truths'', the overall shape of the topics seemingly aligns well with the quark and gluon labels.

Furthermore, the modification plot illustrates a strong match between topic 1 and quark, and topic 2 and gluon, especially in the range where the jet mass histograms have more statistics, namely $[5, 25]$. It does appear that topic 1 and quark match better when the mass is lower, and topic 2 and gluon match better when the mass is higher, but this is likely a byproduct of the number of statistics in the histogram, since the quark peak is to the left of the gluon peak. Nonetheless, the jet mass modification plot seems to demonstrate a good extraction of the quark and gluon values, via the topics, in each bin.

\subsection{Jet Splitting Fraction Extraction}
\begin{figure}[htp]
    \centering
    \begin{subfigure}{0.52\textwidth}
        \centering
        \includegraphics[width=\textwidth]{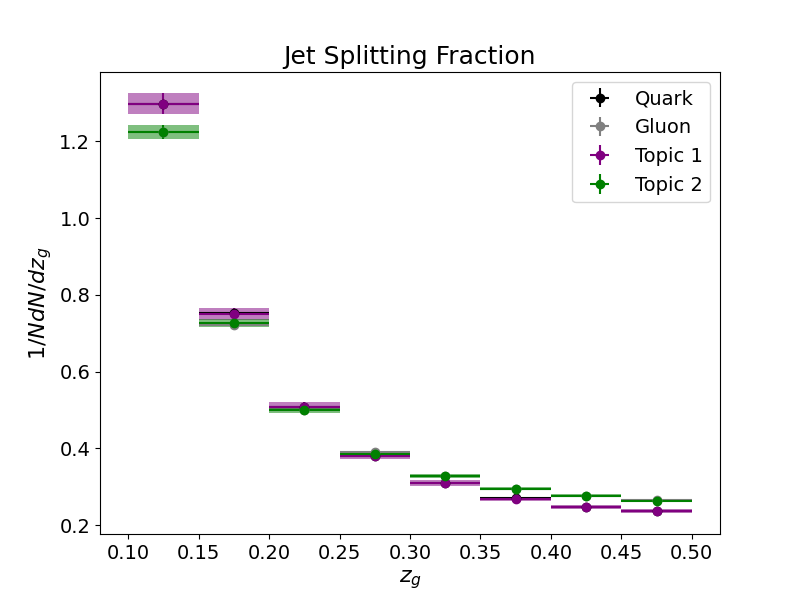}
        \caption{Proton-proton jet splitting function}
        \label{fig:pp-jet-zg}
    \end{subfigure}
    \begin{subfigure}{0.52\textwidth}
        \centering
        \includegraphics[width=\textwidth]{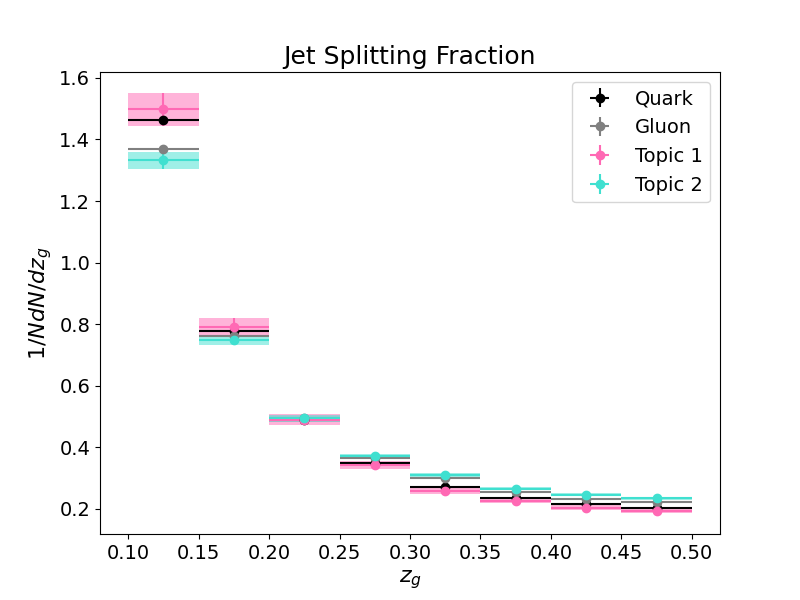}
        \caption{Heavy-ion jet splitting function}
        \label{fig:hi-jet-zg}
    \end{subfigure}
    \begin{subfigure}{0.52\textwidth}
        \centering
        \includegraphics[width=\textwidth]{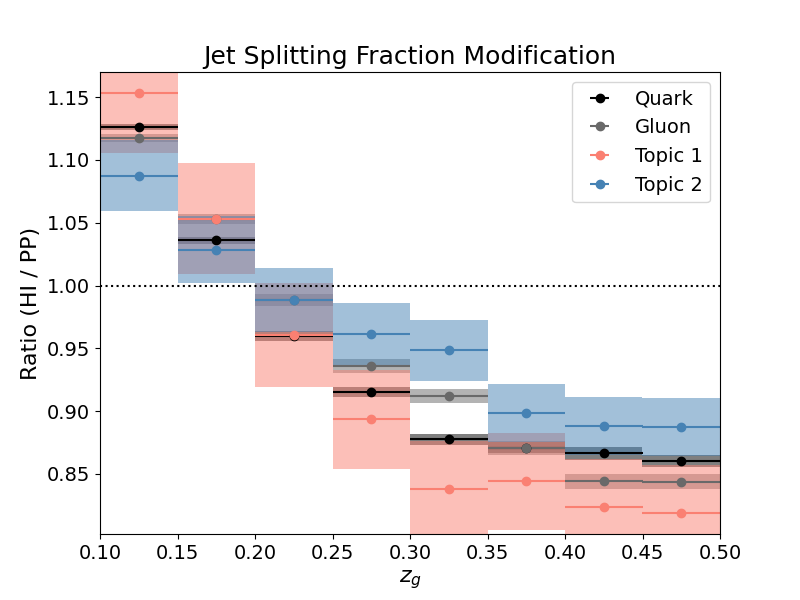}
        \caption{Jet splitting function modification ratio}
        \label{fig:mod-zg}
    \end{subfigure}
    \caption{Topic modeling results for proton-proton (top) and heavy-ion (middle) jet mass, compared to MC truth. Modification of the jet mass in the QGP, as defined by the ratio of the heavy-ion and proton-proton jet mass spectra is shown as well (bottom).}
    \label{fig:jet-zg}
\end{figure}

The jet momentum splitting fraction observable, characterized by $z_g$, describes the momentum ratio of the two leading subjets within the jet. It is defined as $z_g = p_{T,2}/(p_{T,1}+p_{T,2})$, which represents the ratio between the $p_T$ of the subleading subjet over the $p_T$ sum of the two leading subjets \cite{2018-splitting}.

In order to find the subjets, we use SoftDrop \cite{softdrop} / mMDT \cite{mmdt} to decluster the jet's branching history, until transverse momenta of subjets fulfills SoftDrop condition:
\begin{equation}
    \frac{\min{(p_{T,i}, p_{T,j})}}{p_{T,i}+p_{T,j}} > z_{cut}\theta^\beta
\end{equation}
where $\theta$ represents the relative distance between the two subjets. The settings of SoftDrop used for this analysis was $z_{cut} = 0.1$ and $\beta = 0$ \cite{kauder, marzani, 2018-splitting}.

The procedure for extracting the topics is the same as for jet mass. The extracted topics' splitting functions, along with the MC truth quark and gluon splitting functions, are shown in Fig. \ref{fig:jet-zg} for both proton-proton and heavy-ion samples. It appears that the topics match the MC truth labels extremely well for the proton-proton jet splitting fraction, at least well within the uncertainties. However, for the heavy-ion sample, it appears that the MC truth labels are at the edge of or slightly beyond the topics' error bounds.

Thus, the modification plot, measuring the ratio between the PbPb and pp $z_g$ values, appears to perform less well than the previous observables' modification plots. In the previous observables, we observed that the topic 1 corresponds to the quark label and topic 2 corresponds to the gluon label. For the splitting fraction, when we simply compare topic 1 to the quark label or topic 2 to the gluon label, we observe that the topic and the MC truth values are within (or extremely close to) the error bounds for one another.

However, if we compare the topics and the MC truths together, it is not obvious which topic corresponds to which label. Furthermore, comparing the topic 1 and topic 2 splitting fraction modification may result in a different conclusion than comparing the quark and gluon modification, as the topic 1 results relative to topic 2 differ from quark relative to gluon. For example, when $z_g > 0.35$, the topic 2 (supposedly ``gluon'') ratio is greater than the topic 1  (supposedly ``quark'') ratio, while the MC truth labels show the opposite.
\chapter{Machine Learning Observables}

Prior work, as well as results in Section \ref{sec:results}, demonstrate that the constituent multiplicity performs well as an input when extracting topics that are relatively consistent with the quark- and gluon-initiated jet MC truth distributions \cite{brewer}. However, the observables can be further optimized, for example from machine learning techniques, to obtain better separability between the quark and gluon topics \cite{komiske}. Therefore, we explore the possibilities of utilizing supervised learning and classification without labels (CWoLa) to construct new observables that we can input into the topic modeling algorithm.

In this chapter, we introduce a supervised learning technique known as \textit{linear discriminant analysis} (LDA), present a potential feature vector for jets, and give an overview of the training technique used in the presence (or absence) of MC quark/gluon labels. In the next chapter, we will assess the results of the new observable and compare it to the baseline results from topic modeling based on the constituent multiplicity input.

For the purposes of this paper, we choose to use LDA given that it is a technique that can specifically achieve a one-dimensional, linear projection of the data that maximizes separability between the signal and background classes. In addition, LDA requires no hyperparameter tuning, which is convenient as a proof-of-concept method. However, this is not limited to LDA; we could also train a Boosted Decision Tree (BDT) or (deep) neural network to classify between the signal and background, and utilize the (log) probability output as the new observable value.

\section{Linear Discriminant Analysis (LDA)}
Linear discriminant analysis (LDA) is a dimensionality reduction technique that also operates as a linear classifier, separating two or more classes \cite{ghojogh2019linear,Hocker:1019880}. LDA attempts to maximize between-class variance while minimizing within-class variance via a linear projection of the data.

The between-class variance represents the distance between means of different classes:
\begin{equation}
    S_b = \sum_{i=1}^m N_i (\bar{x_i} - \bar{x})(\bar{x_i} - \bar{x})^T
\end{equation}
Here, $m$ represents the total number of classes, $N_i$ represents the number of samples per class, and $x_i$ represents the mean of the samples within the class.

The within-class variance represents the distance between the samples and the mean of a given class:
\begin{equation}
    S_w = \sum_{i=1}^m \sum_{j=1}^{N_i} (x_{i,j} - \bar{x_i})(x_{i,j} - \bar{x_i})^T
\end{equation}

LDA determines the lower-dimensional projection, $P$, such that
\begin{equation}
    P_{LDA} = \argmax_P \frac{|P^T S_b P|}{|P^T S_w P|}
\end{equation}

Once the projection $P$ has been determined, we can transform each feature vector, each representing a jet, in the sample to obtain a single value, which would correspond to the new observable. The distribution of such LDA-projected values determined from the input samples are now the new input distributions into the topic modeling mechanism.

\section{Feature Vector} \label{sec:features}

Here, we introduce the concept of \textbf{ML multiplicity}. A jet's constituent multiplicity is simply the count of constituent particles within a jet cone. However, a jet's track multiplicity within certain $p_T^{track}$ bins is known to be a good quark/gluon discriminant. Therefore, we can decompose this multiplicity into an $n$-dimensional vector, where $n$ represents some number of $p_T^{track}$ bins and each value in the vector represents the number of constituents with a $p_T^{track}$ within the bin range corresponding to that vector element index.

\begin{figure}
    \centering
    \centering
    \includegraphics[width=.5\textwidth]{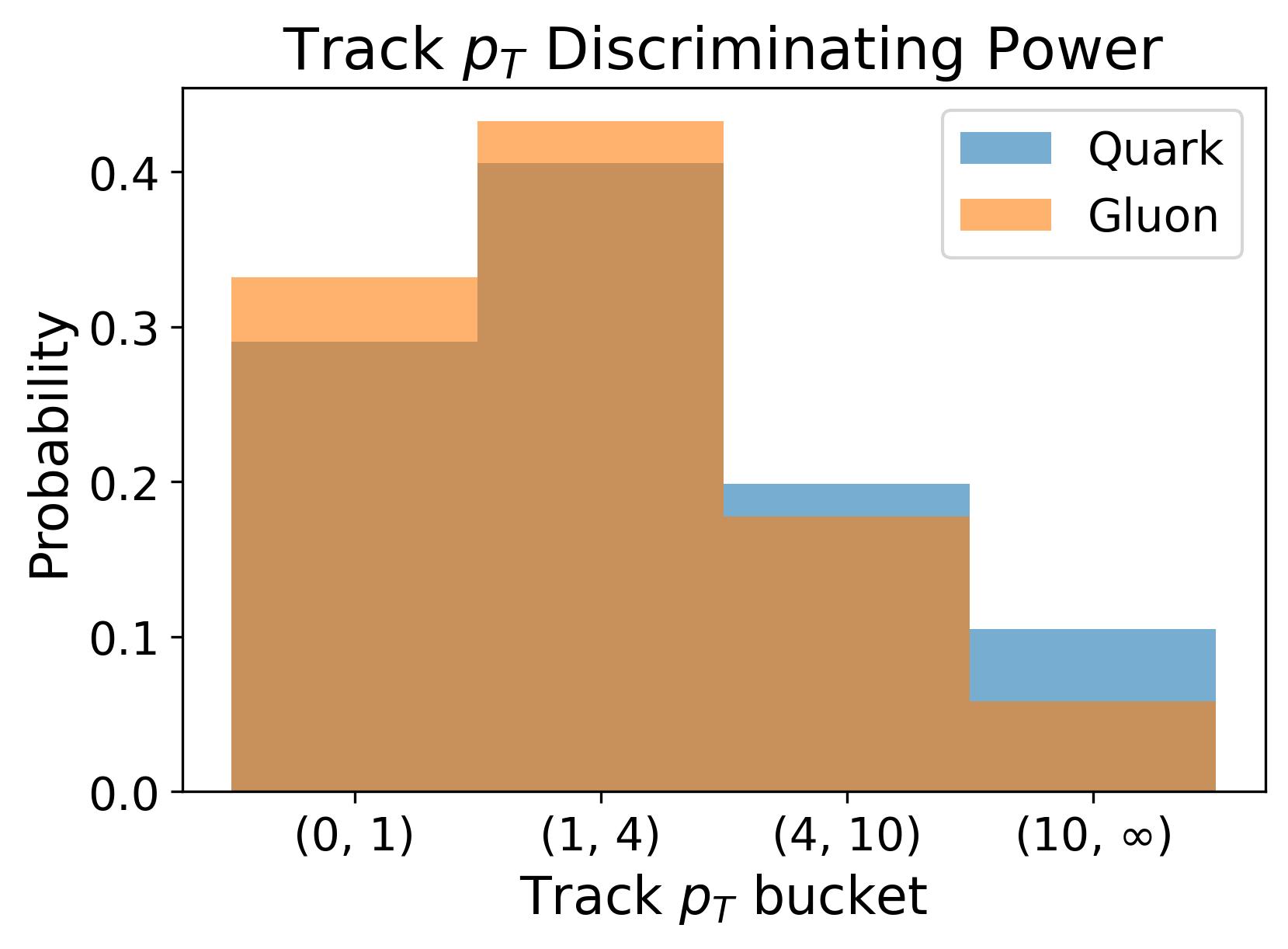}
    \caption{Probability density of constituents in track $p_T$ buckets for quark-like and gluon-like jets}
    \label{fig:trackpt_disc}
\end{figure}

In particular, we create a 4-dimensional vector to represent each jet, $[x_1, x_2, x_3, x_4]$, where $x_i$ corresponds to the following:
\begin{center}
\begin{tabular}{ c l } \label{tab:ml-mult}
 $x_1$: & \# of constituents with $0 < p_T^{track} \leq 1$ GeV \\  
 $x_2$: & \# of constituents with $1 < p_T^{track} \leq 4$ GeV \\  
 $x_3$: & \# of constituents with $4 < p_T^{track} \leq 10$ GeV \\  
 $x_4$: & \# of constituents with $p_T^{track} > 10$ GeV
\end{tabular}
\end{center}
The sum, $x_1 + x_2 + x_3 + x_4$, should be equal to the constituent multiplicity of the jet. Fig. \ref{fig:trackpt_disc} shows the probability density of tracks in each $p_T^{track}$ bin for both quark and gluon jets, with respect to the MC label, illustrating that this feature vector may be appropriate to achieve better quark/gluon discrimination.

This is the feature vector we will feed into the supervised learning algorithm to achieve a projection that we can then use for topic modeling. Since this paper focuses on LDA, we will term the multiplicity projection as the \textbf{LDA multiplicity}.

However, it is also worth mentioning that there are other observables that may seem promising as well, and potentially even better to use in a realistic collider scenario. Prior work has proven that count observables (such as constitutent multiplicity) performs better than shape observables with respect to the MC generator truth \cite{komiske}. However, count observables are also often difficult to accurately measure in realistic settings and generally not IRC-safe, meaning that any minuscule perturbation will change the result. In these situations, relying purely on count observables renders the extension of this algorithm to more realistic data questionable. Other observables that are also known to be good quark/gluon discriminators are $p_T^D$ \cite{higgssearch}, LHA (``Les Houches Angularity'' \cite{badger2016les}), width (related to jet broadening), and mass (related to jet thrust) \cite{badger2016les, qg-tagging, metodiev-cwola}.

\section{Machine Learning Methodology}
Previous work has demonstrated as a proof-of-concept that using machine learning methods to derive an observable can enhance the separation power between quark and gluon jets, and rely less on count observables \cite{komiske, metodiev-cwola}. This section will discuss different ways to leverage machine learning, namely supervised learning and classification without labels, to craft new observables.

\subsection{Quark/Gluon Supervised Learning}
Since we are attempting to separate quark- and gluon-like jets, the most straightforward and obvious supervised learning exercise is to simply train a classifier to discriminate between MC-labeled quark jets and MC-labeled gluon jets. Thus, we are leveraging the fact that we have access to the Monte Carlo ``truth'' labels to train a classifier. The approach we use is to train the discriminator on proton-proton samples, and apply the resulting model to heavy-ion inputs.

\subsection{Classification Without Labels (CWoLa)} \label{sec:cwola}
The downside of the aforementioned technique in the previous section is that it requires quark/gluon truth labels in the data, which we do not have access to in realistic collider scenarios. However, a previously introduced paradigm known as \textit{classification without labels}, or CWoLa, allows classifiers to be trained directly on data in scenarios where labels may be unknown, such as the given one \cite{metodiev-cwola}. Under CWoLa, we simply need to fulfill the assumptions that the training samples are pure (uncontaminated) mixed samples. The theorem that describes the principle behind CWoLa is as follows:
\\\\
\textbf{CWoLa Theorem.} \cite{metodiev-cwola} \textit{Given mixed samples $A$ and $B$ defined in terms of pure sample $1$ and $2$ with signal fractions $f_A > f_B$ in the equations below, an optimal classifier trained to distinguish $A$ from $B$ is also optimal for distinguishing 1 from 2}
\begin{align*}
p_{A}(\vec{x}) = f_A p_{1}(\vec{x}) + (1-f_A)p_{2}(\vec{x}),  \\
p_{A}(\vec{x}) = f_B p_{1}(\vec{x}) + (1-f_B)p_{2}(\vec{x})
\end{align*}

The pure sample 1 and pure sample 2 distributions in this theorem correspond to the base distributions (topics) in the topic modeling algorithm, and by extension, the quark- and gluon-like jet distributions. The mixed samples are the input dijet and $\gamma+$jet samples, which we will assume fulfills the CWoLa assumption of being uncontaminated mixtures of quark- and gluon-initiated jets.

The CWoLa paradigm argues that an optimal classifier, $h$, trained using full supervision to discriminate between the input samples, is the same optimal classifier to distinguish between the base distributions. Ref. \cite{metodiev-cwola} proves this via demonstrating that the likelihood ratio $L_{A / B} (\vec{x}) = p_{A}/p_{B}$ is a monotonically increasing rescaling of the likelihood ratio $L_{S1/S2}$ if $f_A > f_B$ (if $f_A < f_B$, then classifier is reversed).

Though the optimal classifier remains the same, the operating point $c$, which can be interpreted as the classification output threshold value, of the classifier $h$ changes. While it is possible to directly classify between the signal and background samples if we know the value of $c$, this requires additional knowledge, such some sample where the mixture proportions, $f_{A}$ and $f_{B}$, are known \cite{metodiev-cwola}.

However, without such a sample, we can still leverage the CWoLa paradigm by combining it with topic modeling, such as in Ref. \cite{komiske}. In order to do so, we train a model on dijet and $\gamma+$jet samples, given the feature vectors described in \ref{sec:features}. The output of the model is then fed into the topic modeling machinery.

\section{LDA Application}

In order to perform the LDA transformation, we use the TMVA (Toolkit for Multivariate Data Analysis) \cite{Hocker:2007ht} library in ROOT. To form the training set, we construct a dataset consisting of feature vectors that contain 50\% of the $\gamma+$jet sample and 50\% of the dijet sample. Each jet in the training set is randomly selected from the respective input sample.

Once we train the classifier, we apply the LDA transformation to the entire input sample for both $\gamma+$jet and dijets. Though traditional machine learning advise against utilizing the training data for purposes other than training, throwing away such samples would be quite wasteful.\footnote{We tried various combinations here: using the entire sample for training and transformation, using part of the sample for training and the rest for transformation, using part of the sample for training and the entire sample for transformation, etc. In practice, there has not been an issue with using the full dataset for the projection.}

In addition, it is worth noting that the test set is somewhat irrelevant in our scenario, because we have not defined an operating point for our classifier. This means that we are not utilizing LDA for direct classification, and thus the classification accuracy does not matter much. If comparisons amongst various types of classifiers (such as BDTs and neural nets) are desired, the receiver operating characteristic (ROC) and the area under the ROC curve (AUC) are better summaries than classification accuracy.

\chapter{Machine Learning Observables: Results}

\section{Quark/Gluon Discrimination} \label{sec:mult-lda}
\begin{figure*}[htp]
    \centering
    \begin{subfigure}{.47\textwidth}
        \centering
        \includegraphics[width=\textwidth]{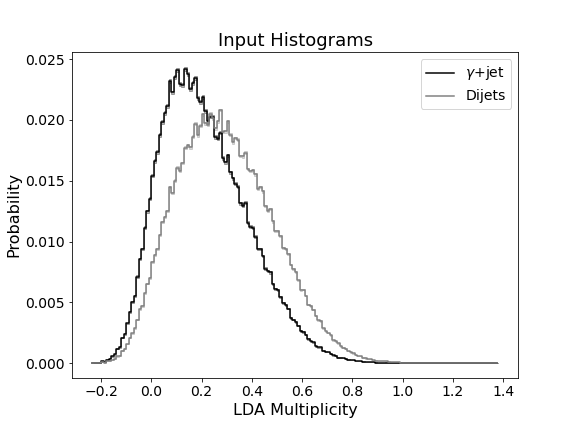}
        \caption{Proton-proton input distributions}
        \label{fig:pp-input-lda}
    \end{subfigure}
    \begin{subfigure}{.47\textwidth}
        \centering
        \includegraphics[width=\textwidth]{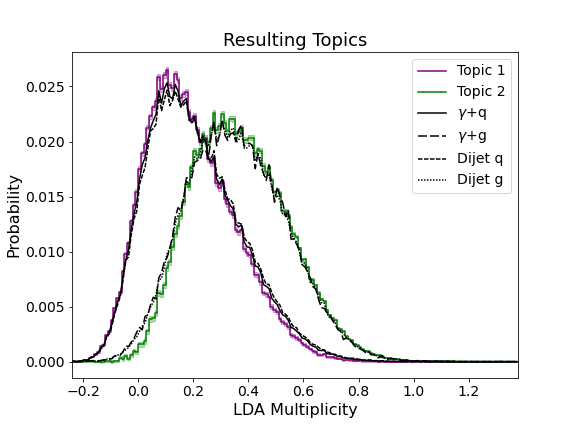}
        \caption{Proton-proton topic results}
        \label{fig:pp-res-lda}
    \end{subfigure}
    \begin{subfigure}{.47\textwidth}
        \centering
        \includegraphics[width=\textwidth]{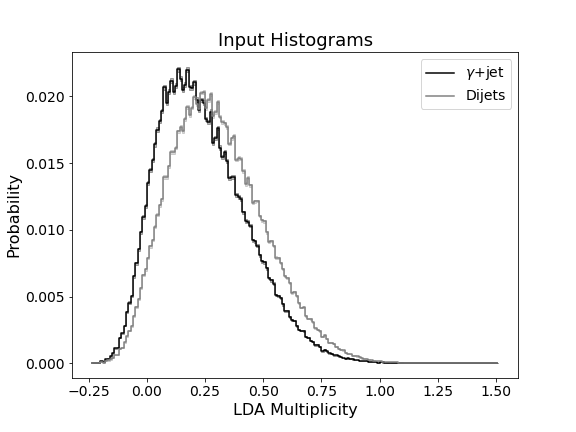}
        \caption{Heavy-ion input distributions}
        \label{fig:pbpb-input-lda}
    \end{subfigure}
    \begin{subfigure}{.47\textwidth}
        \centering
        \includegraphics[width=\textwidth]{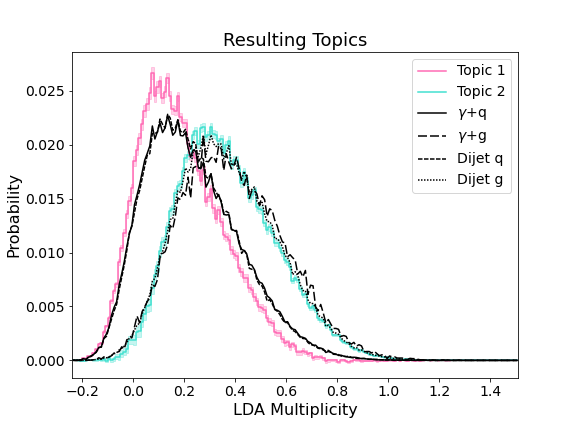}
        \caption{Heavy-ion topic results}
        \label{fig:pbpb-res-lda}
    \end{subfigure}
    \caption{The input density distributions and corresponding topic results for pp (top) and PbPb (bottom) samples, where the $x$-axis represents the LDA projection of the jets' track $p_T$ feature vector trained to discriminate between quark and gluon}
    \label{fig:mult-lda}
\end{figure*}

The first approach is to train a classifier to directly discriminate between quark and gluon jets, using the Monte Carlo simulation labels, on pp data, then apply the model to PbPb data. The motivation behind training on pp data is because simply that pp collision simulations are more robust and reliable. Supposing that any QGP modification is second order effect, the quark/gluon classification should be captured by pp simulation. Thus, it is more reasonable to start from a better simulation for training the separation model, then apply the trained model to the PbPb sample. Topic modeling has also been already been applied experimentally and compared to simulation on pp collisions at ATLAS \cite{atlas}. From this approach, we can obtain the input histogram of the transformed observable values for $\gamma+$jet and dijet samples for both proton-proton and heavy-ion collisions, as shown on the left in Fig. \ref{fig:mult-lda}.

The resulting topics from the LDA-projected histograms are shown on the right in Fig. \ref{fig:mult-lda}. Comparing these results to the constituent multiplicity results in Fig. \ref{fig:topics}, they are pretty consistent, with the exception of the heavy-ion topic 2 appearing to matching slightly better with the MC truth label for the gluon distributions. Therefore, it appears that including some machine learning to craft another observable can, in fact, help improve the performance of the topic modeling, though the effect may be fairly minimal.

Furthermore, Appendix \ref{app:a} contains figures that demonstrate the extraction of jet substructure plots for both pp and PbPb, as well as the ratio between heavy-ion and proton-proton jet substructure observables. It appears that for LDA multiplicity trained to distinguish between quark and gluon, the use of supervised learning does not seem to affect the value of each bin much, but rather simply increase the uncertainty.

\section{CWoLa}

Given that in a realistic setting, it is impossible to classify individual jets as quark or gluon jets, so training a classifier to distinguish between them is out of question. However, we can use the CWoLa paradigm mentioned in Section \ref{sec:cwola}. The advantage using CWoLa is that we can train a separate discriminator for the pp and the PbPb samples, and we have absolute knowledge of the labels used during the classification, making this approach purely data-driven.

\begin{figure*}[htp]
    \centering
    \begin{subfigure}{.47\textwidth}
        \centering
        \includegraphics[width=\textwidth]{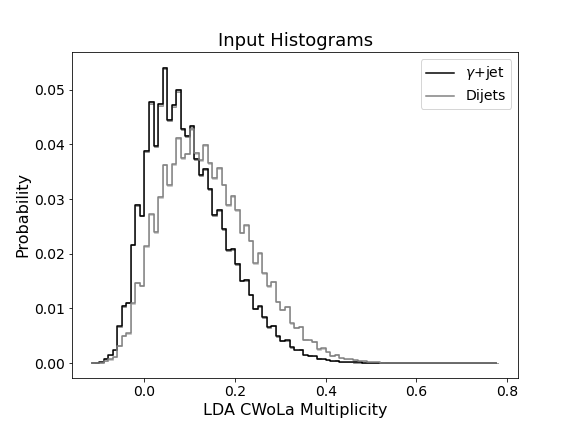}
        \caption{Proton-proton input distributions}
        \label{fig:pp-in-cwola}
    \end{subfigure}
    \begin{subfigure}{.47\textwidth}
        \centering
        \includegraphics[width=\textwidth]{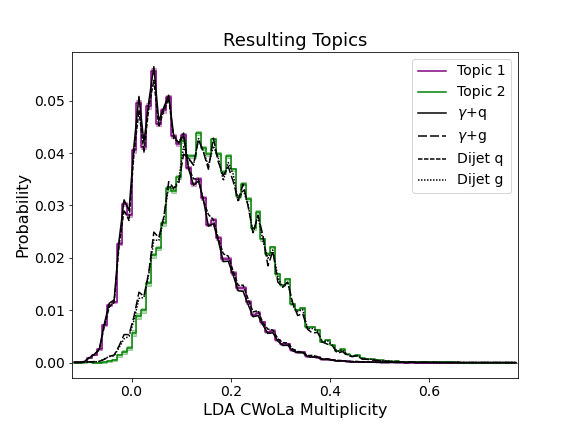}
        \caption{Proton-proton topic results}
        \label{fig:pp-out-cwola}
    \end{subfigure}
    \begin{subfigure}{.47\textwidth}
        \centering
        \includegraphics[width=\textwidth]{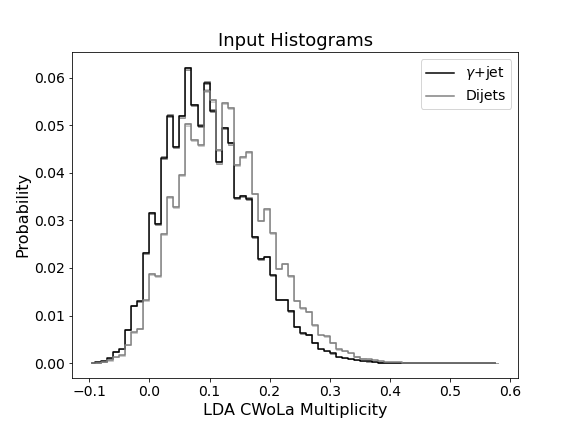}
        \caption{Heavy-ion input distributions}
        \label{fig:pbpb-in-cwola}
    \end{subfigure}
    \begin{subfigure}{.47\textwidth}
        \centering
        \includegraphics[width=\textwidth]{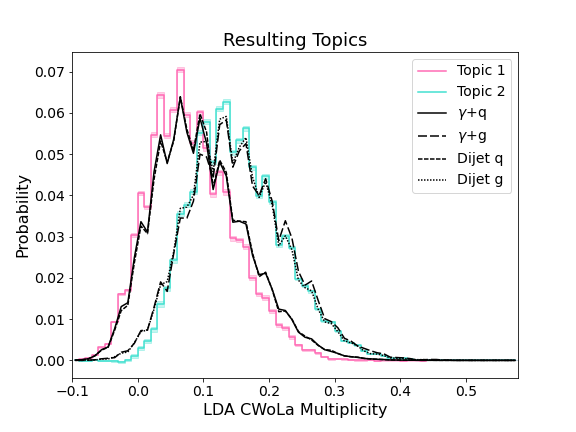}
        \caption{Heavy-ion topic results}
        \label{fig:pbpb-out-cwola}
    \end{subfigure}
    \caption{The input distributions and output topics for pp (top) and PbPb (bottom) samples. The $x$-axis represents the LDA projection of the jets' track $p_T$ feature vector using the CWoLA paradigm (training to discriminate between the input samples)}
    \label{fig:mult-cwola}
\end{figure*}

The input LDA-projected distributions and the resulting topics are shown in Fig. \ref{fig:mult-cwola}. While it's not obvious that there is any increased separability between the $\gamma+$jet and dijet samples from the input histograms, the resulting topics do appear to align better with the MC truth labels.

In the proton-proton sample, topic 1 appears to almost match exactly with the quark truth, and topic 2 appears to match quite well with the gluon truth. In the heavy-ion sample, it appears that the results of topic modeling outperform both the constituent multiplicity results and the LDA results from Sec. \ref{sec:mult-lda}. Topic 1 appears to ``fill out'' the top of the quark distribution better than in previous experiment and topic 2 appears to fit the gluon truth extremely well.

In the figures in Appendix \ref{app:a}, it appears that the $p_T^{track}$ multiplicity trained using CWoLa results show potential for improvement over the constituent multiplicity baseline results. In particular, the pp topic 1 results are noticeably closer to the quark truth labels for all four substructure observables (jet shape, jet fragmentation, jet mass, and jet splitting fraction), while the pp topic 2 results and the PbPb results appear unchanged.

\section{Comparing Fractions} \label{sec:compare-results}

While we only need the $\kappa$ values in order to calculate the quark/gluon jet spectra and substructures, we can also compute the fraction of each base distribution in the input samples and compare the results, as another indicator of topic modeling performance. Recall Eq. \ref{eq:mix} represents the input distributions as mixtures of some base distributions. Rewriting this equation for clarity, we can express the $\gamma+$jet and dijet distributions as:
\begin{equation}
    \begin{split}
        p_{\gamma+jet} = f_A b_1(x) + (1-f_A) b_2(x),\\
        p_{dijet} = f_B b_1(x) + (1-f_B) b_2(x)
    \end{split}
\end{equation}
Thus, $f_A$ represents the fraction of topic 1 in the $\gamma+$jet sample, and $1-f_A$ represents the fraction of topic 2 in the $\gamma+$jet sample. Similarly, $f_B$ represents the fraction of topic 1 in the dijet sample, and $1-f_B$, topic 2.

The $\kappa$ values are related to the fractions via the following equations \cite{brewer}:
\begin{equation}
    \kappa_{AB} = \frac{1-f_A}{1-f_B}, \ \ 
    \kappa_{BA} = \frac{f_B}{f_A} 
\end{equation}

We can derive the expression for $f_A$ and $f_B$ from the $\kappa$ values in Eq. \ref{eq:topics}:
\begin{equation}
    \begin{split}
        f_A = \frac{1-\kappa_{AB}}{1-\kappa_{AB}\kappa_{BA}},\\
        f_B = \frac{\kappa_{BA}(1-\kappa_{AB})}{1-\kappa_{AB}\kappa_{BA}}
    \end{split}
\end{equation}

\begin{figure*}[tp]
    \centering
    \begin{subfigure}{.47\textwidth}
        \centering
        \includegraphics[width=\textwidth]{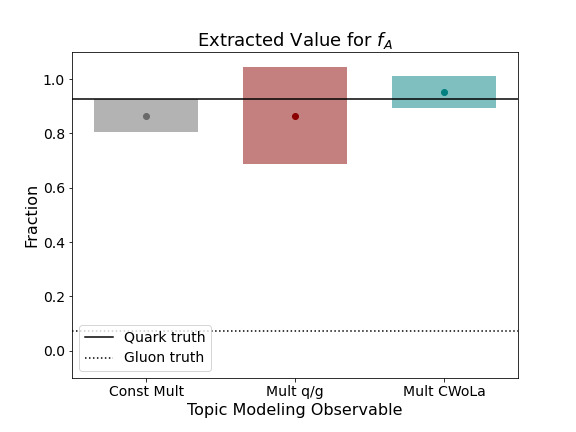}
        \caption{pp $\gamma+$jet sample}
        \label{fig:ppf1}
    \end{subfigure}
    \begin{subfigure}{.47\textwidth}
        \centering
        \includegraphics[width=\textwidth]{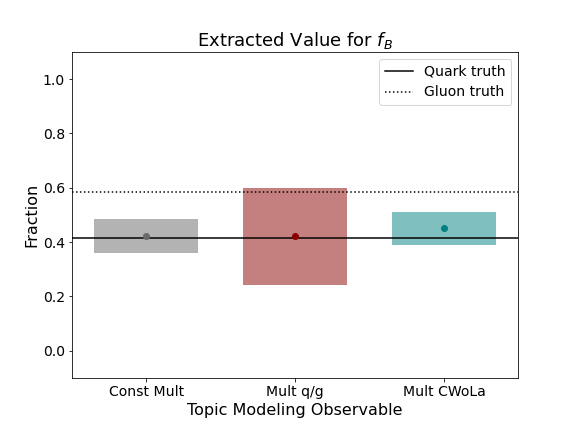}
        \caption{pp dijet sample}
        \label{fig:ppf2}
    \end{subfigure}
    \begin{subfigure}{.47\textwidth}
        \centering
        \includegraphics[width=\textwidth]{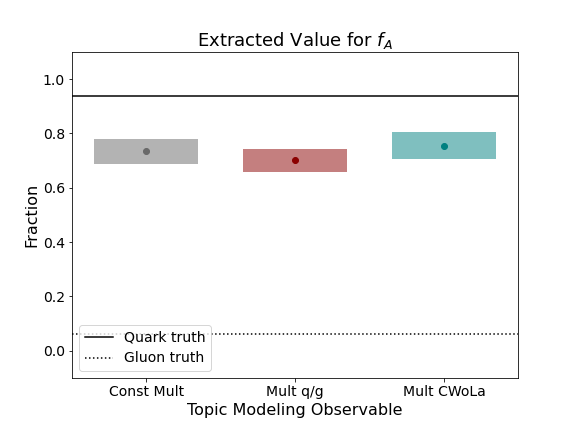}
        \caption{PbPb $\gamma+$jet sample}
        \label{fig:pbpbf1}
    \end{subfigure}
    \begin{subfigure}{.47\textwidth}
        \centering
        \includegraphics[width=\textwidth]{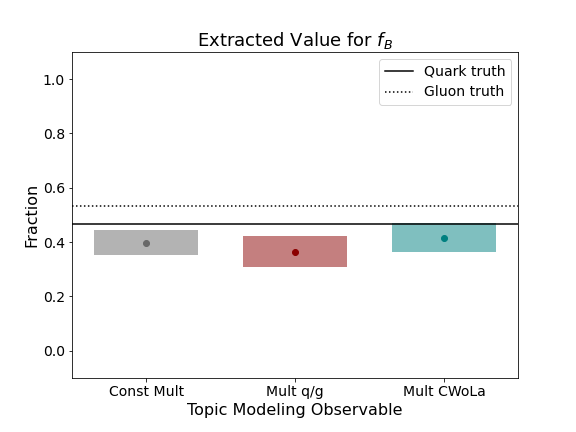}
        \caption{PbPb dijet sample}
        \label{fig:pbpbf2}
    \end{subfigure}
    \caption{Extracted value of $f_A$ and $f_B$ for pp and PbPb samples, compared to MC quark and gluon truth values. Here, $f_A$ represents the fraction of topic 1 jets in sample A ($\gamma+$jet in this case) and $1-f_A$ represents the fraction of topic 2 jets in the same sample. $f_B$ and $1-f_B$ represent the fractions of topic 1 jets and topic 2 jets, respectively, in sample B (dijets in this case).}
    \label{fig:compare}
\end{figure*}

Therefore, in order to have a better sense of the ``accuracy'' of each approach, relative to the true value, we plot the extracted topic 1 fraction values next to one another in Fig. \ref{fig:compare}, for each input sample. In addition, we also plot the estimated quark and gluon fractions in the PYQUEN samples, keeping in mind that $f_{gluon} = 1-f_{quark}$. While the Monte Carlo ``truth'' values are the closest values we have to the ultimate truth, this is a gentle reminder that they are not to be taken as the absolute true. Recall, we match jets to the ``truth'' particle based on the nearest matrix element, as determined by $\Delta R$. In addition, we only account for jets in the sample where the outgoing matrix element is $\Delta R < 0.4$ from the jet axis, which is not all the jets in the sample, as stated in Section \ref{sec:data}.

In Fig. \ref{fig:compare}, we once again see that topic 1 aligns well with the quark truth value. For the proton-proton sample, we notice that the constituent multiplicity and the quark/gluon-trained result in approximately the same value, but the supervised learning approach has a larger error, similar to what was observed in the substructure observables extraction.

The CWoLa-trained result obtains a different value for the fraction, but one that seemingly overestimates the quark fraction in both the $\gamma+$jet and dijet samples. In the $\gamma+$jet sample, we observe that the CWoLa-trained result appears slightly closer to the truth, while in the dijet sample, we observe that the CWoLa-trained result appears slightly further from the MC truth. The slight improvement in the $\gamma+$jet is consistent with the analysis for the extracted jet substructure observables in Appendix \ref{app:a}, and it demonstrates that there is \textit{potential} in creating an observable that improves performance of the algorithm. Nonetheless, in all three scenarios, the values are quite close to one another and to the quark truth, all within uncertainty.

In the heavy-ion case, we observe that in both the $\gamma+$jet and the dijet samples, we underestimate the quark sample in all the input scenarios. It is worth noting that this quark underestimation is a phenomenon also consistent with Brewer's observations using topic modeling on a JEWEL dataset \cite{brewer}.

It appears that in both input samples, the quark/gluon training actually results in a value further from the MC quark truth, while the CWoLa training results in a value closer to the quark truth. For the quark/gluon-trained observable, this is no surprise, as the training had been done on a proton-proton sample, and simply applied to the heavy-ion sample. This could be evidence that perhaps transfer learning for a new observable is an area that could be explored. Nonetheless, these results appear consistent with the substructure observable results in Appendix \ref{app:a}, as well. In the dijet sample, the quark truth is within the uncertainty range for the CWoLa-trained value. Once again, overall, all the results indicate that there is potential for topic modeling improvement using machine learning.
\chapter{Discussion/Conclusion}

In summary, we have corroborated previous proof-of-concept results for fully data-driven jet topic separation, using Monte Carlo samples from the PYQUEN generator. The resulting topics from the topic modeling machinery appear to be in good agreement with the MC generator level truths for quark-like and gluon-like distributions, though there appears to be slight disagreements, especially when extracting the quark and gluon jet substructure observables. One possible explanation for these disparities may be that constituent multiplicity is not an optimal input into the topic modeling algorithm, and additional observables, such as ones derived from machine learning, may be required to yield better results \cite{komiske}.

On the other hand, one assumption that is made in this algorithm is that the resulting mutually irreducible base distributions correspond to the quark- and gluon-like jets. Therefore, another potential reason for the observed differences may be simply that this assumption does not hold. In other words, the mutually irreducible base distributions are simply not associated with the quark- and gluon-initiated jet distributions. Alternatively, it may be possible that the quark- and gluon-like jets in dijets are slightly different from the quark- and gluon-like jets in the $\gamma+$jets, which violates the assumption that the signal and background distributions are the same in the mixed input samples. Practically, one could still extract the mutually irreducible jet topics using the data-driven technique and compare with the same extraction in the theoretical calculation or predictions from event generators to work around this problem.

Another issue is that there are ambiguities in defining the per-jet MC truth label, as we define the truth labels to only include jets that are within $\Delta R = \sqrt{\Delta \eta^2 + \Delta \phi^2} = 0.4$ from a quark or gluon outgoing matrix element, which does not include all jets in the sample. This implies that our input samples of the $\gamma+$jet and dijets, from which we extract topic 1 and topic 2, are not pure mixtures of quark and gluon jets, as we have defined them. Perhaps even more unsettling, the labels of ``quark'' and ``gluon'' themselves are fundamentally ambiguous for reasons mentioned in Ref. \cite{qg-tagging}.

One observation that Brewer makes is that the topic modeling algorithm in JEWEL consistently finds a larger gluon-like fraction compared to the MC labels, which is consistent in our PYQUEN results as well \cite{brewer}. This may be attributed to a quark-initiated jet becoming more gluon-like through gluon radiation, meaning it may be possible that there are changes to jet structure during the collision that may lead them to appear either more quark-like or gluon-like.

Nonetheless, our results from PYQUEN-generated Monte Carlo samples corroborate previous proof-of-concept studies performed using JEWEL, demonstrating that a fully data-driven technique can potentially be used to extract separate quark and gluon jet distributions from experimental samples, without additional pieces of knowledge. We further extended the proof-of-concept study by demonstrating that resulting fractions can be applied to $\gamma+$jet and dijet observable values to extract jet substructure observables, such as jet shape and jet fragmentation. The results also show potential for extracting quark and gluon per-jet substructure spectra, involving per-jet quantities, such as jet mass and jet splitting fraction. While there are still disparities between the topics and the MC-labeled quark- and gluon-like values, these results suggest potential for an experimental determination of quark and gluon jet spectra and their substructure.

The substructure observables obtained from the topics can then be used to approximate the quark and gluon jet substructure modification in the quark-gluon plasma. Despite these aforementioned discrepancies, we are ultimately able to extract the jet substructure modification by observing the ratio between heavy-ion and proton-proton jet substructures. While the uncertainty is quite large and sometimes the topic does not correspond well to the true quark/gluon value, the modification plots demonstrate that approximate ratio values can be extracted.

Since these deviations may demonstrate limitations of the input observable, we explore the potential of additional observables, constructed from machine learning, specifically focusing on LDA. The LDA multiplicity observable, constructed from feature vectors representing multiplicity in various $p_T^{track}$ bins, appears to improve the discrimination power between the quark and gluon topics. The supervised learning results illustrate potential for greater quark/gluon mutual irreducibility from an input observable created from machine learning.

With regards to future work, in addition to training models using multiplicity in various $p_T^{track}$ bins, there are other discriminating observables, such as jet $p_T^D$, LHA, width, and mass, that hold quark/gluon separating potential \cite{badger2016les, qg-tagging, metodiev-cwola}. We hope to further analyze these observables and utilize them effectively in training additional machine learning models. Furthermore, there are limitations with the linear nature of the LDA algorithm. We hope to perform the same analysis using nonlinear models, such as boosted decision trees and (deep) neural nets, which may allow for improved quark/gluon discrimination. We encourage further exploration of observables and models that may ultimately increase the quark/gluon separability of input samples, improving the results of the topic modeling algorithm.
\appendix
\chapter{Code Availability}

The code for this topic modeling analysis can be found at
\href{https://github.com/kying18/jet-topics}{https://github.com/kying18/jet-topics}.
\chapter{Figures} \label{app:a}

\begin{figure*}[htp]
    \centering
    \begin{subfigure}{.48\textwidth}
        \centering
        \includegraphics[width=\textwidth]{plots/plots_redo/pp-shape.png}
        \caption{pp multiplicity result}
         
    \end{subfigure}
    \begin{subfigure}{.48\textwidth}
        \centering
        \includegraphics[width=\textwidth]{plots/plots_redo/pbpb-shape.png}
        \caption{PbPb multiplicity result}
         
    \end{subfigure}
    \begin{subfigure}{.48\textwidth}
        \centering
        \includegraphics[width=\textwidth]{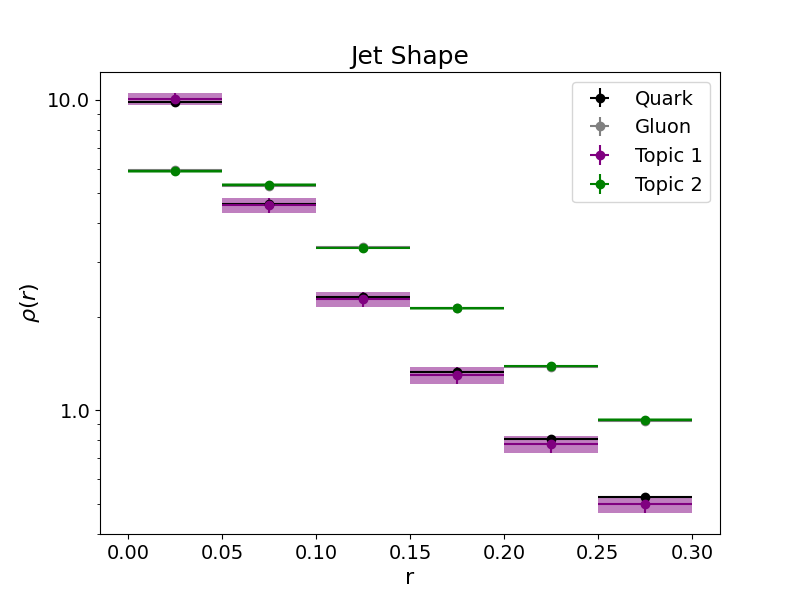}
        \caption{pp LDA multiplicity result}
         
    \end{subfigure}
    \begin{subfigure}{.48\textwidth}
        \centering
        \includegraphics[width=\textwidth]{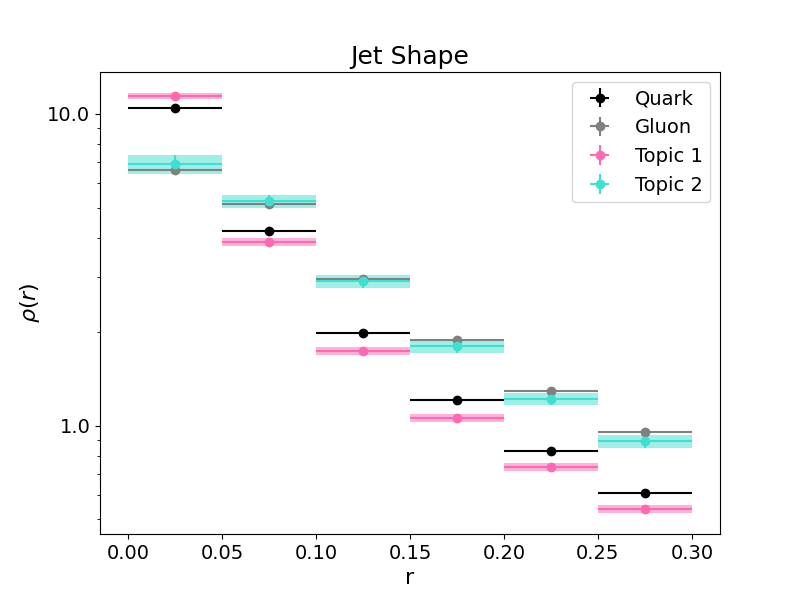}
        \caption{PbPb LDA multiplicity result}
         
    \end{subfigure}
    \centering
    \begin{subfigure}{.48\textwidth}
        \centering
        \includegraphics[width=\textwidth]{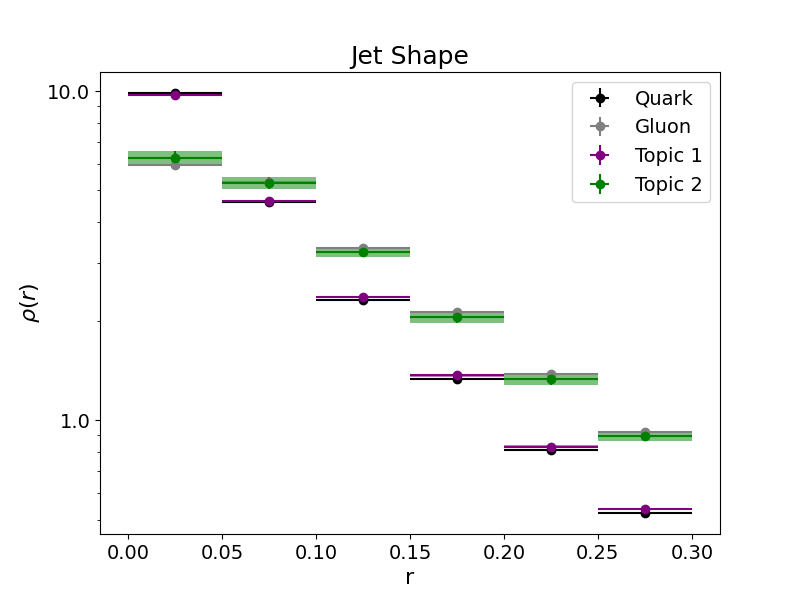}
        \caption{pp multiplicity CWoLa result}
         
    \end{subfigure}
    \begin{subfigure}{.48\textwidth}
        \centering
        \includegraphics[width=\textwidth]{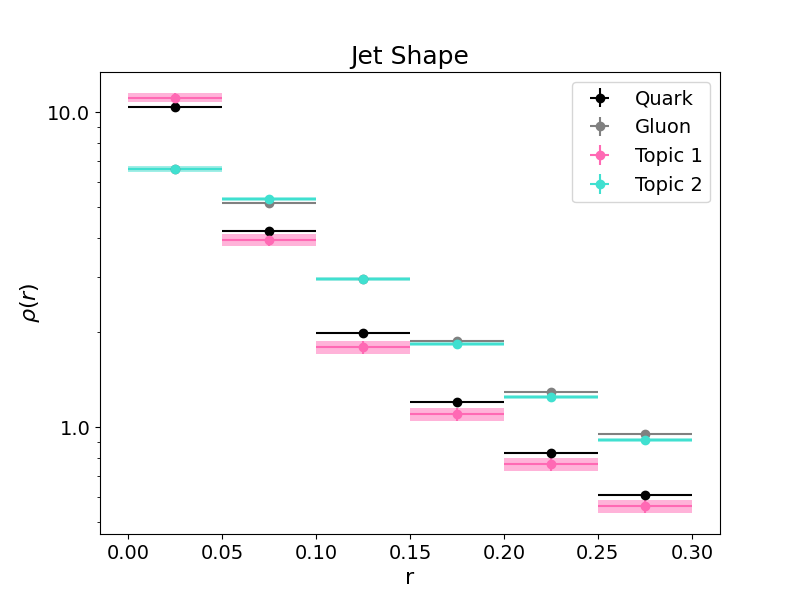}
        \caption{PbPb multiplicity CWoLa result}
         
    \end{subfigure}
    %
    %
    %
    %
    
    \caption{Jet shape extraction result for pp (left) and PbPb (right) for the constituent multiplicity topic modeling baseline, followed by topic modeling results performed on supervised learning inputs}
    \label{fig:ml-jet-shape}
\end{figure*}
\clearpage
\newpage

\begin{figure*}[htp]
    \centering
    \begin{subfigure}{.5\textwidth}
        \centering
        \includegraphics[width=\textwidth]{plots/plots_redo/mod-shape.png}
        \caption{Constituent multiplicity result}
         
    \end{subfigure}
    \begin{subfigure}{.5\textwidth}
        \centering
        \includegraphics[width=\textwidth]{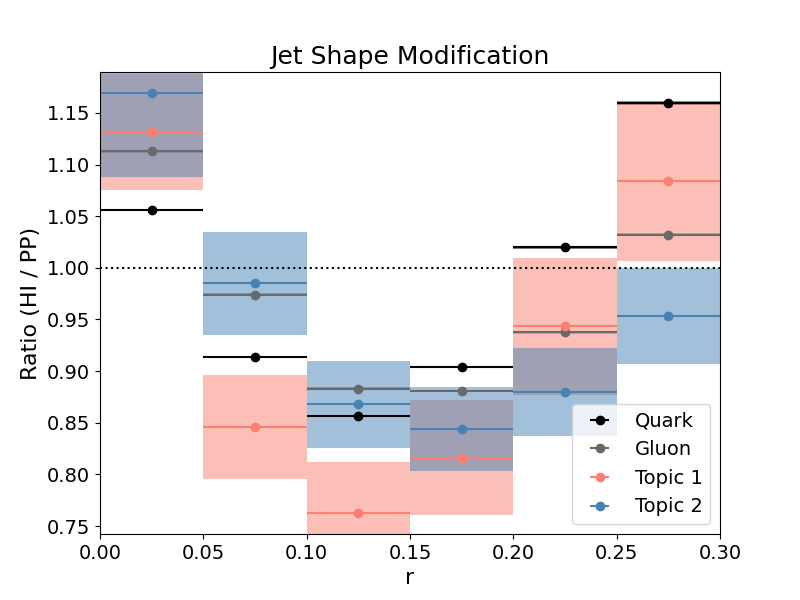}
        \caption{LDA multiplicity result}
         
    \end{subfigure}
    \begin{subfigure}{.5\textwidth}
        \centering
        \includegraphics[width=\textwidth]{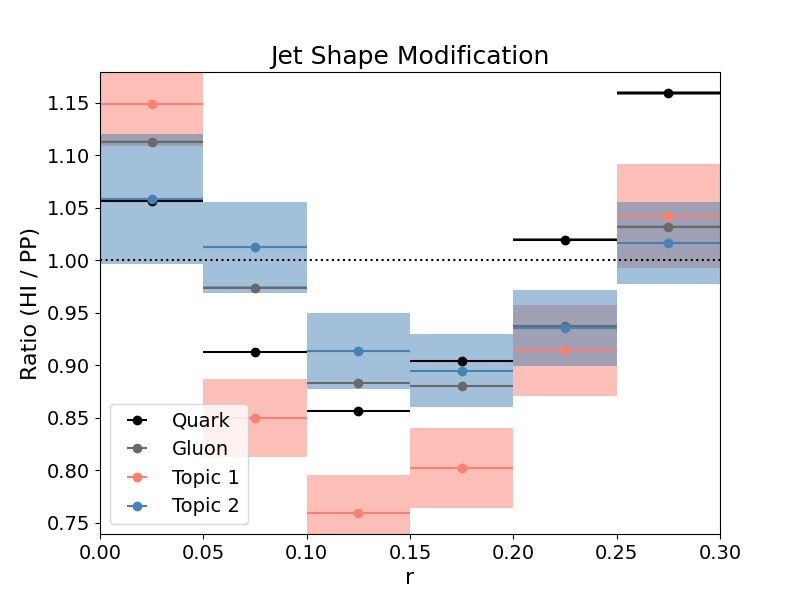}
        \caption{Multiplicity CWoLa result}
         
    \end{subfigure}
    %
    %
    
    \caption{Jet shape modification results for the constituent multiplicity topic modeling baseline, followed by topic modeling results performed on supervised learning inputs}
    \label{fig:ml-jet-shape-mod}
\end{figure*}
\clearpage
\newpage

\begin{figure*}[htp]
    \centering
    \begin{subfigure}{.48\textwidth}
        \centering
        \includegraphics[width=\textwidth]{plots/plots_redo/pp_settings_100,35000,30000_jet-frag.png}
        \caption{pp multiplicity result}
    \end{subfigure}
    \begin{subfigure}{.48\textwidth}
        \centering
        \includegraphics[width=\textwidth]{plots/plots_redo/settings_100,35000,30000_jet-frag.png}
        \caption{PbPb multiplicity result}
    \end{subfigure}
    \begin{subfigure}{.48\textwidth}
        \centering
        \includegraphics[width=\textwidth]{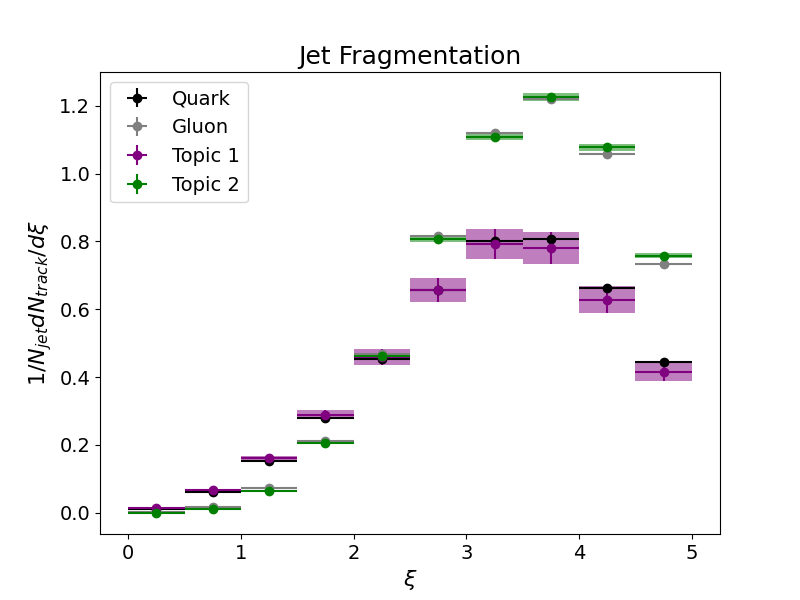}
        \caption{pp LDA multiplicity result}
    \end{subfigure}
    \begin{subfigure}{.48\textwidth}
        \centering
        \includegraphics[width=\textwidth]{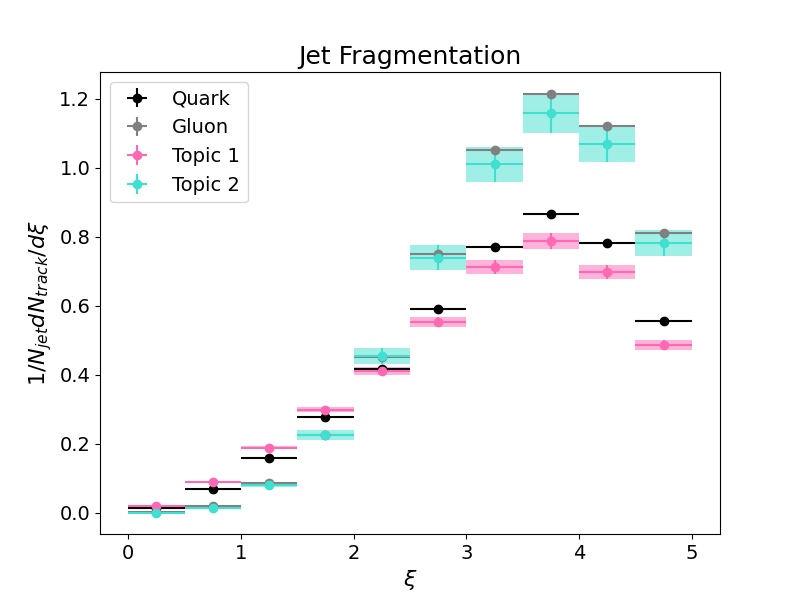}
        \caption{PbPb LDA multiplicity result}
    \end{subfigure}
    \begin{subfigure}{.48\textwidth}
        \centering
        \includegraphics[width=\textwidth]{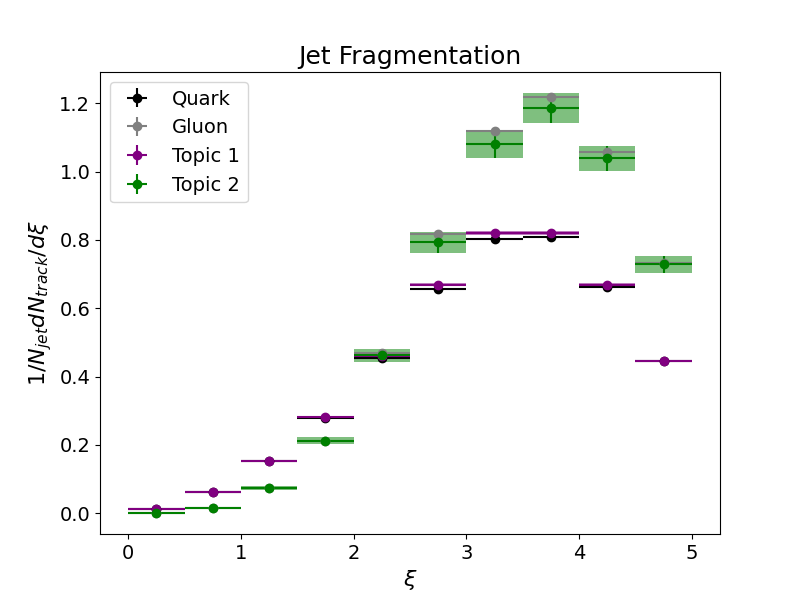}
        \caption{pp multiplicity CWoLa result}
    \end{subfigure}
    \begin{subfigure}{.48\textwidth}
        \centering
        \includegraphics[width=\textwidth]{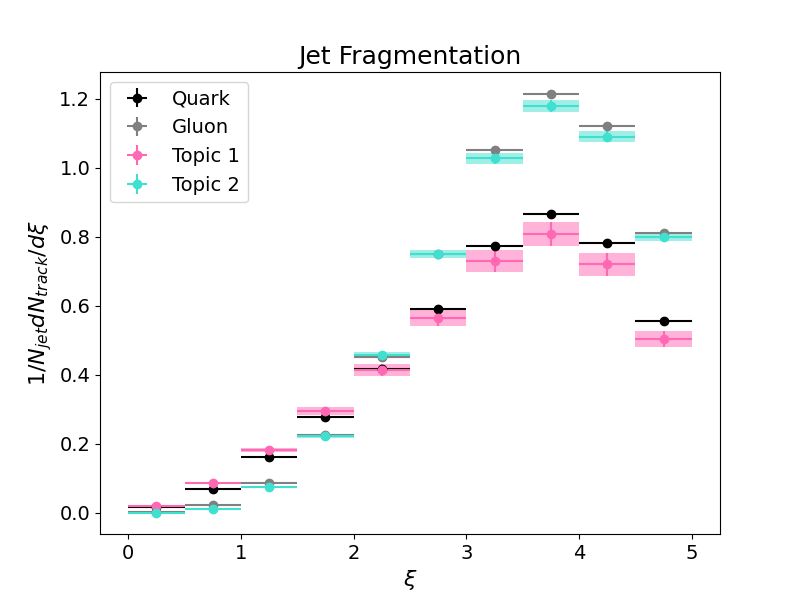}
        \caption{PbPb multiplicity CWoLa result}
    \end{subfigure}
%
%
    
    \caption{Jet fragmentation extraction result for pp (left) and PbPb (right) for the constituent multiplicity topic modeling baseline, followed by topic modeling results performed on supervised learning inputs}
    \label{fig:ml-jet-frag}
\end{figure*}
\clearpage
\newpage

\begin{figure*}[htp]
    \centering
    \begin{subfigure}{.5\textwidth}
        \centering
        \includegraphics[width=\textwidth]{plots/plots_redo/mod-settings_100,35000,30000_jet-frag.png}
        \caption{Constituent multiplicity result}
         
    \end{subfigure}
    \begin{subfigure}{.5\textwidth}
        \centering
        \includegraphics[width=\textwidth]{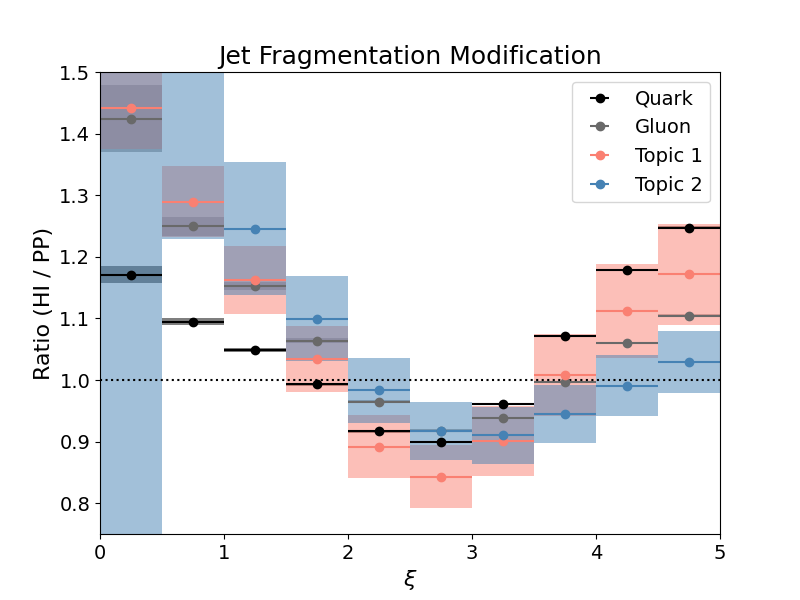}
        \caption{LDA multiplicity result}
         
    \end{subfigure}
    \begin{subfigure}{.5\textwidth}
        \centering
        \includegraphics[width=\textwidth]{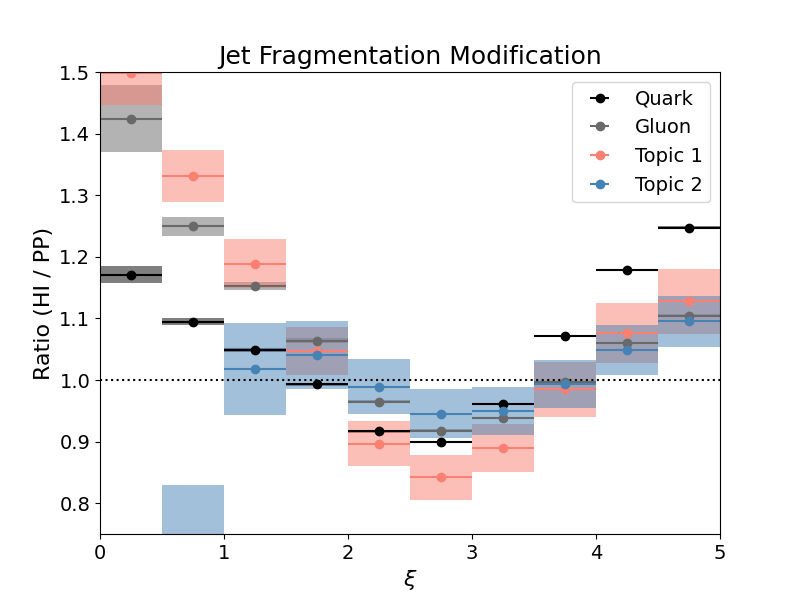}
        \caption{Multiplicity CWoLa result}
         
    \end{subfigure}
    %
    %
    
    \caption{Jet fragmentation modification results for the constituent multiplicity topic modeling baseline, followed by topic modeling results performed on supervised learning inputs}
    \label{fig:ml-jet-frag-mod}
\end{figure*}
\clearpage
\newpage

\begin{figure*}[htp]
    \centering
    \begin{subfigure}{.48\textwidth}
        \centering
        \includegraphics[width=\textwidth]{plots/plots_redo/pp-mass.png}
        \caption{pp multiplicity result}
         
    \end{subfigure}
    \begin{subfigure}{.48\textwidth}
        \centering
        \includegraphics[width=\textwidth]{plots/plots_redo/pbpb-mass.png}
        \caption{PbPb multiplicity result}
         
    \end{subfigure}
    \begin{subfigure}{.48\textwidth}
        \centering
        \includegraphics[width=\textwidth]{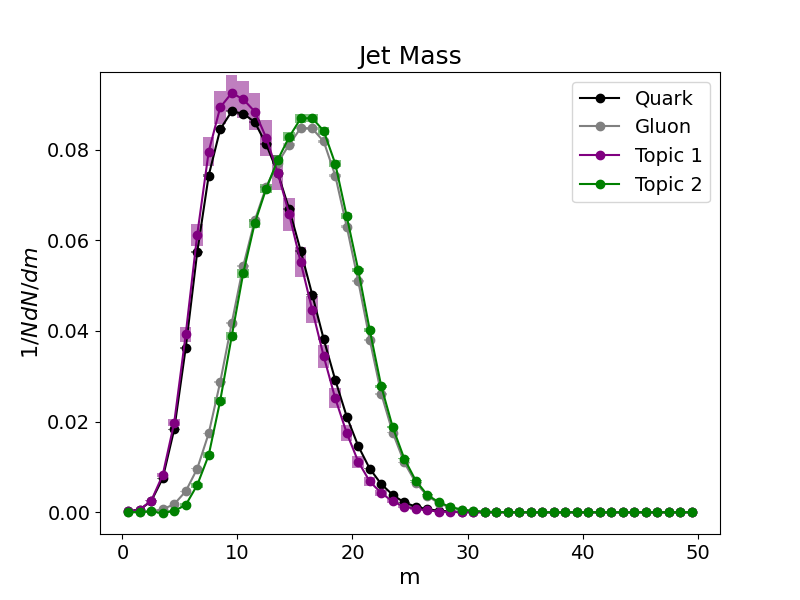}
        \caption{pp LDA multiplicity result}
         
    \end{subfigure}
    \begin{subfigure}{.48\textwidth}
        \centering
        \includegraphics[width=\textwidth]{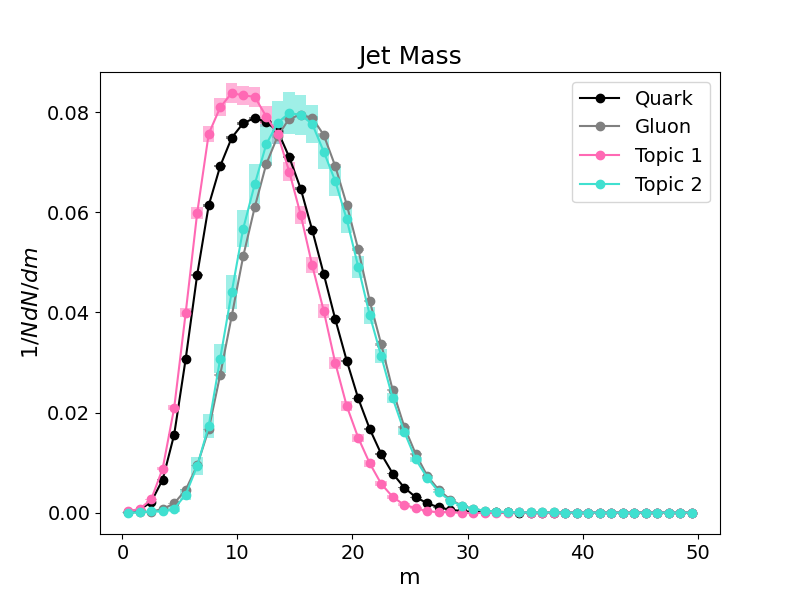}
        \caption{PbPb LDA multiplicity result}
         
    \end{subfigure}
    \begin{subfigure}{.48\textwidth}
        \centering
        \includegraphics[width=\textwidth]{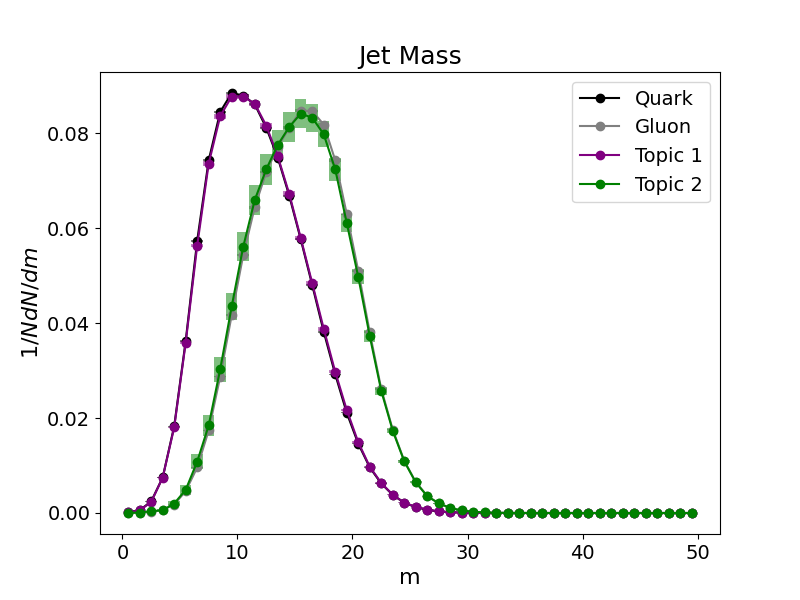}
        \caption{pp multiplicity CWoLa result}
         
    \end{subfigure}
    \begin{subfigure}{.48\textwidth}
        \centering
        \includegraphics[width=\textwidth]{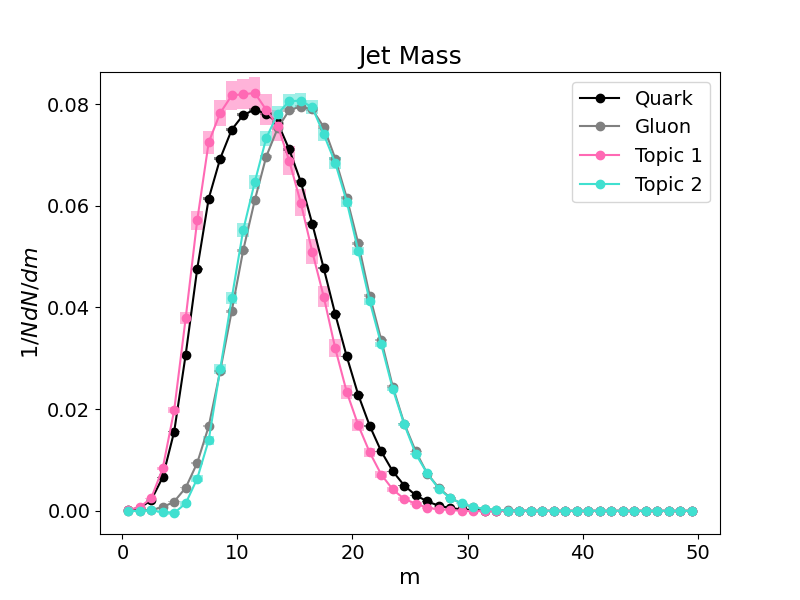}
        \caption{PbPb multiplicity CWoLa result}
         
    \end{subfigure}
%
%
%
%
    
    \caption{Jet mass extraction result for pp (left) and PbPb (right) for the constituent multiplicity topic modeling baseline, followed by topic modeling results performed on supervised learning inputs}
    \label{fig:ml-jet-mass}
\end{figure*}
\clearpage
\newpage

\begin{figure*}[htp]
    \centering
    \begin{subfigure}{.5\textwidth}
        \centering
        \includegraphics[width=\textwidth]{plots/plots_redo/mod-mass.png}
        \caption{Constituent multiplicity result}
         
    \end{subfigure}
    \begin{subfigure}{.5\textwidth}
        \centering
        \includegraphics[width=\textwidth]{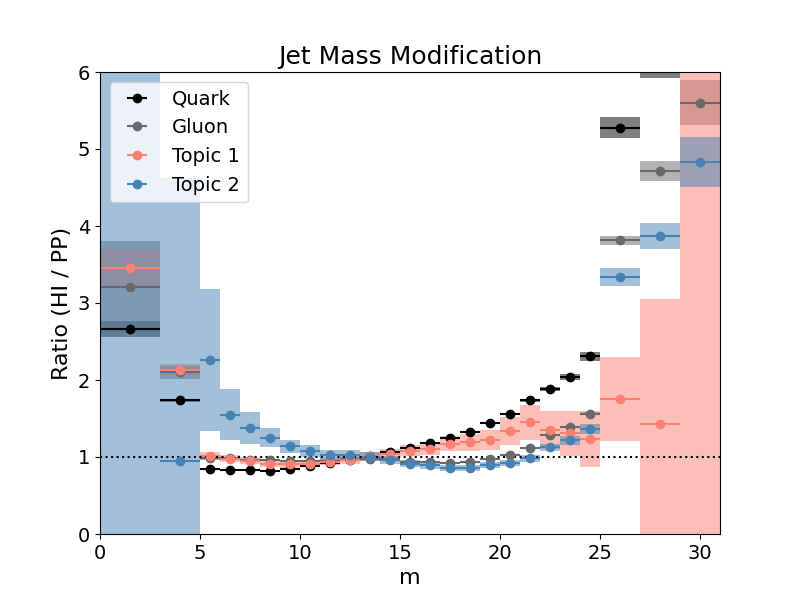}
        \caption{LDA multiplicity result}
         
    \end{subfigure}
    \begin{subfigure}{.5\textwidth}
        \centering
        \includegraphics[width=\textwidth]{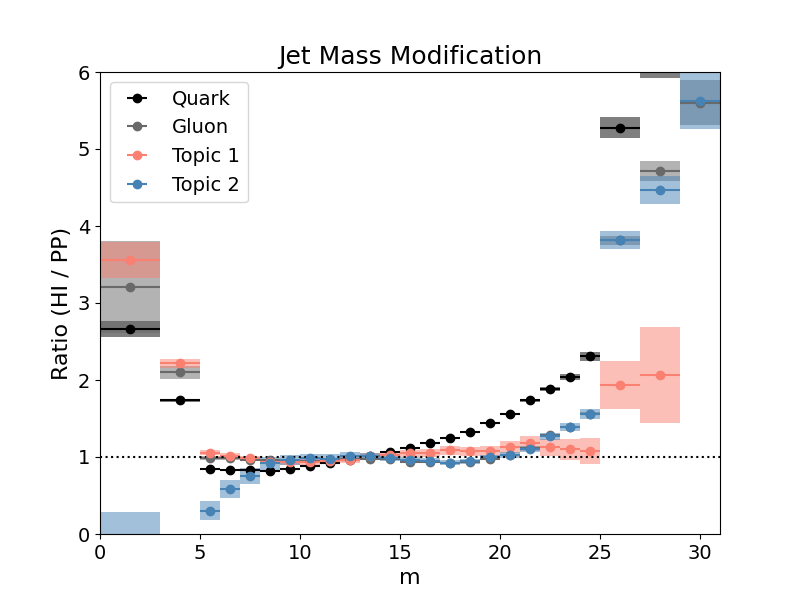}
        \caption{Multiplicity CWoLa result}
         
    \end{subfigure}
    %
    %
    
    \caption{Jet mass modification results for the constituent multiplicity topic modeling baseline, followed by topic modeling results performed on supervised learning inputs}
    \label{fig:ml-jet-mass-mod}
\end{figure*}
\clearpage
\newpage

\begin{figure*}[htp]
    \centering
    \begin{subfigure}{.48\textwidth}
        \centering
        \includegraphics[width=\textwidth]{plots/plots_redo/pp-splitting.png}
        \caption{pp multiplicity result}
         
    \end{subfigure}
    \begin{subfigure}{.48\textwidth}
        \centering
        \includegraphics[width=\textwidth]{plots/plots_redo/pbpb-splitting.png}
        \caption{PbPb multiplicity result}
         
    \end{subfigure}
    \begin{subfigure}{.48\textwidth}
        \centering
        \includegraphics[width=\textwidth]{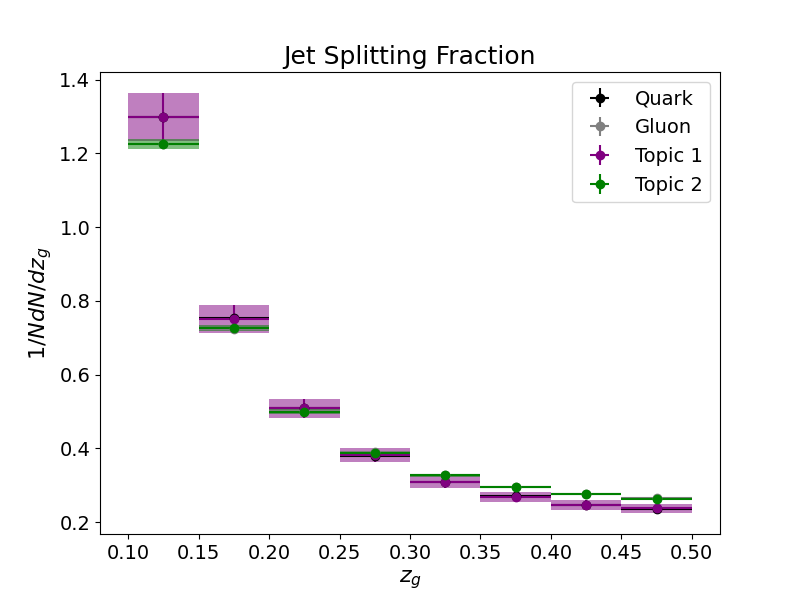}
        \caption{pp LDA multiplicity result}
         
    \end{subfigure}
    \begin{subfigure}{.48\textwidth}
        \centering
        \includegraphics[width=\textwidth]{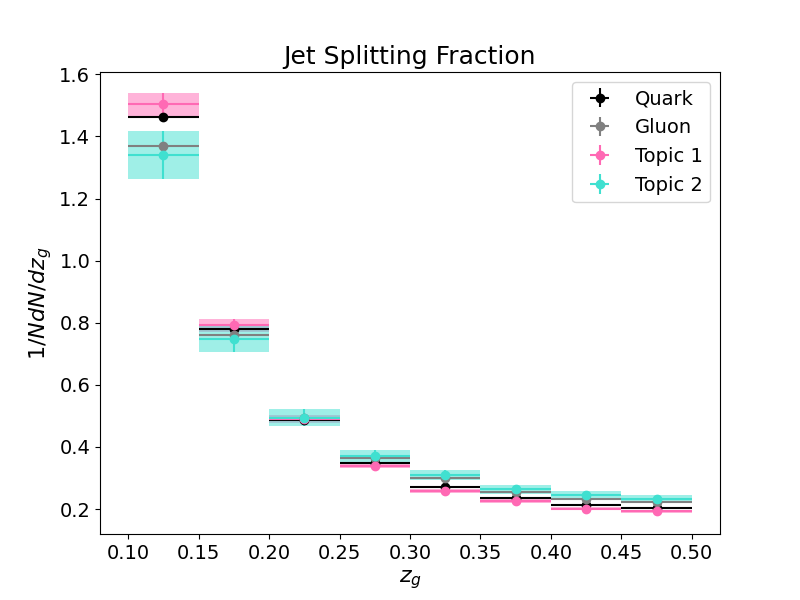}
        \caption{PbPb LDA multiplicity result}
         
    \end{subfigure}
    \begin{subfigure}{.48\textwidth}
        \centering
        \includegraphics[width=\textwidth]{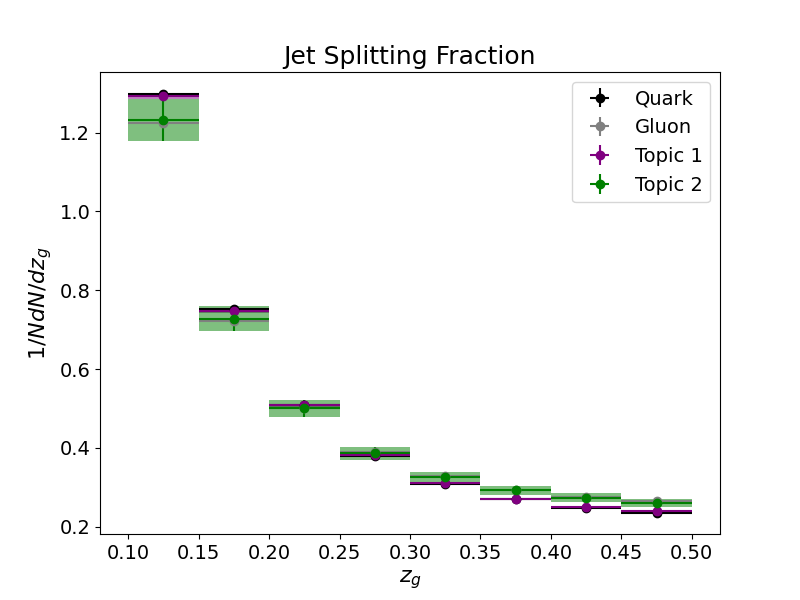}
        \caption{pp multiplicity CWoLa result}
         
    \end{subfigure}
    \begin{subfigure}{.48\textwidth}
        \centering
        \includegraphics[width=\textwidth]{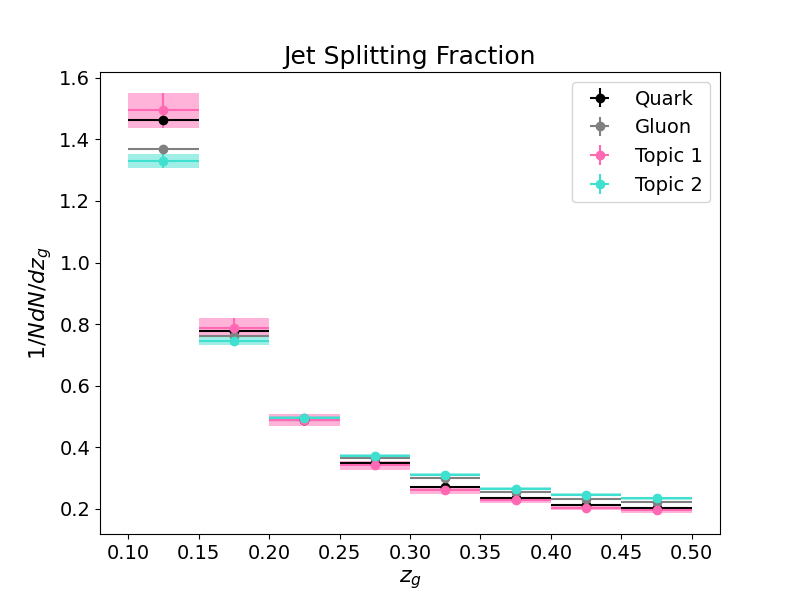}
        \caption{PbPb multiplicity CWoLa result}
         
    \end{subfigure}
%
%
%
%
    
    \caption{Jet splitting fraction extraction result for pp (left) and PbPb (right) for the constituent multiplicity topic modeling baseline, followed by topic modeling results performed on supervised learning inputs}
    \label{fig:ml-jet-zg}
\end{figure*}
\clearpage
\newpage

\begin{figure*}[htp]
    \centering
    \begin{subfigure}{.5\textwidth}
        \centering
        \includegraphics[width=\textwidth]{plots/plots_redo/mod-splitting.png}
        \caption{Constituent multiplicity result}
         
    \end{subfigure}
    \begin{subfigure}{.5\textwidth}
        \centering
        \includegraphics[width=\textwidth]{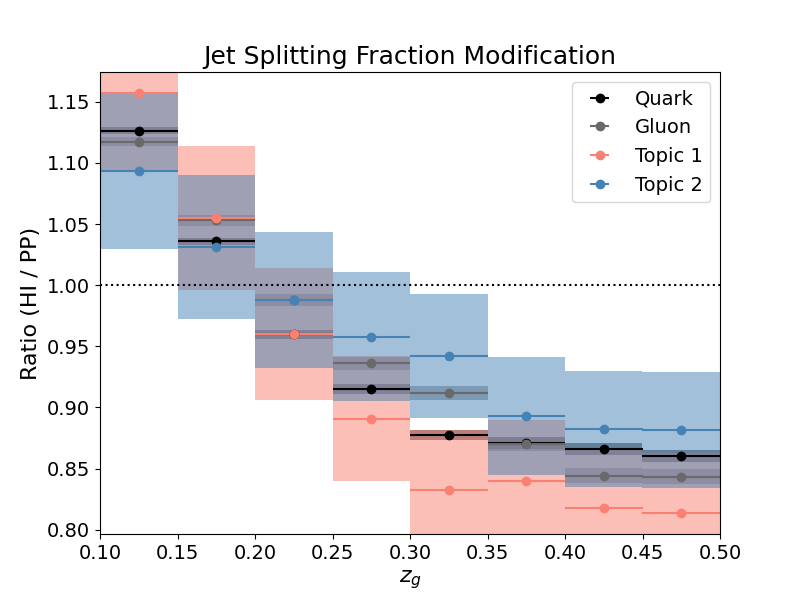}
        \caption{LDA multiplicity result}
         
    \end{subfigure}
    \begin{subfigure}{.5\textwidth}
        \centering
        \includegraphics[width=\textwidth]{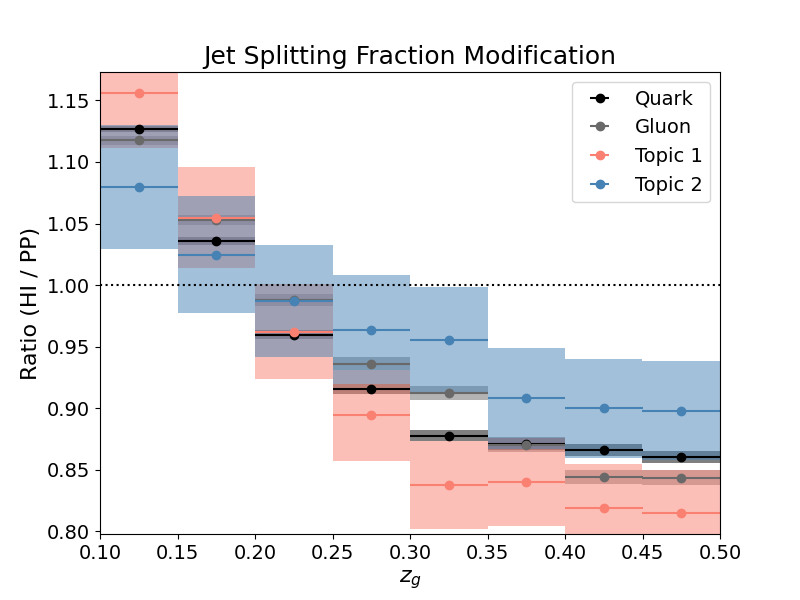}
        \caption{Multiplicity CWoLa result}
         
    \end{subfigure}
    %
    %
    
    \caption{Jet splitting fraction modification results for the constituent multiplicity topic modeling baseline, followed by topic modeling results performed on supervised learning inputs}
    \label{fig:ml-jet-splitting-mod}
\end{figure*}
\clearpage
\newpage
\begin{singlespace}
\bibliography{main}
\bibliographystyle{unsrt}
\end{singlespace}
\end{document}